\documentclass[12pt,preprint]{aastex}

\shorttitle{Orbital Motion in Pre-Main Sequence Binaries}
\shortauthors{Schaefer et al.}

\bibpunct[]{(}{)}{;}{a}{}{,}

\begin{document}

\title{Orbital Motion in Pre-Main Sequence Binaries}

\author{G. H. Schaefer\altaffilmark{1}, L. Prato\altaffilmark{2}, M. Simon\altaffilmark{3}, and J. Patience\altaffilmark{4,5}}

\altaffiltext{1}{The CHARA Array of Georgia State University, Mount Wilson Observatory, Mount Wilson, CA 91023, U.S.A. (schaefer@chara-array.org)}
\altaffiltext{2}{Lowell Observatory, 1400 West Mars Hill Road, Flagstaff, AZ 86001, USA}
\altaffiltext{3}{Department of Physics and Astronomy, Stony Brook University, Stony Brook, NY 11794, USA}
\altaffiltext{4}{Astrophysics Group, School of Physics, University of Exeter, Exeter, EX4 4QL, UK}
\altaffiltext{5}{School of Earth and Space Exploration, Arizona State University, PO Box 871404, Tempe, Arizona, 85287, USA}

\begin{abstract}

We present results from our ongoing program to map the visual orbits of pre-main sequence binaries in the Taurus star forming region using adaptive optics imaging at the Keck Observatory.  We combine our results with measurements reported in the literature to analyze the orbital motion for each binary.  We present preliminary orbits for DF Tau, T Tau S, ZZ Tau, and the Pleiades binary HBC 351.  Seven additional binaries show curvature in their relative motion.  Currently, we can place lower limits on the orbital periods for these systems; full solutions will be possible with more orbital coverage.  Five other binaries show motion that is indistinguishable from linear motion.  We suspect that these systems are bound and might show curvature with additional measurements in the future.  The observations reported herein lay critical groundwork toward the goal of measuring precise masses for low-mass pre-main sequence stars.

\end{abstract}

\keywords{binaries: visual, stars: pre-main sequence, stars: fundamental parameters}

\section{Introduction}

The mass and chemical composition of a star determines how its physical properties evolve over time.  At pre-main sequence (PMS) ages, there are significant differences in the evolutionary tracks predicted by different sets of models \citep[e.g.,][]{hillenbrand04,mathieu07,simon08}.  Mapping the orbit of a binary star provides a way to measure the dynamical masses of the stellar components and test predictions of the evolutionary models.  

At the distance to the Taurus star forming region \citep[140 pc;][]{kenyon94}, a binary with a total mass of $1~M_\odot$ and a period of 10 yr would be separated by only 33 milli-arcseconds (mas) on the sky.  Therefore, binaries in nearby star forming regions that can be resolved through high resolution techniques such as speckle interferometry or adaptive optics imaging, typically have periods longer than a decade \citep[e.g.,][]{schaefer06}.  At these long periods, orbital variations in the radial velocities of the components are small and can be difficult to measure \citep[e.g.,][]{nguyen12,schaefer12}.  If a binary orbit can be resolved spatially and spectroscopically, then the individual masses of the components and the distance to the system can be measured. If only the visual orbit is known, then the total mass can help provide a constraint on the evolutionary tracks.  Additionally, distributions of the orbital parameters (period, semi-major axis, eccentricity) can be used to test theories of the formation and dynamical evolution of binary stars \citep{mathieu94,goodwin07}.  Long-baseline optical/infrared interferometry has the power to resolve shorter period double-lined spectroscopic binaries in nearby star forming regions \citep{boden05,boden07,schaefer08,simon13}.

In this paper, we present results from an ongoing program to map the visual orbits of binaries in the Taurus star forming region using adaptive optics imaging at the Keck Observatory.  The sample of binaries has separations between 30--300 mas.  For several of the shorter period binaries (separation $\lesssim$ 100 mas), we obtained observations on a yearly basis.  We inserted longer period systems (separations between 100--300 mas) into our observing program as time allowed.  Combined with previous astrometric measurements reported in the literature, we plotted the orbital motion observed for each system over time.  We computed preliminary orbital solutions for four systems, found that another seven show curvature in their relative motion but the measurements do not yet provide a reliable set of orbital parameters, and that five additional binaries have limited orbital coverage that is currently indistinguishable from linear motion.  We discuss each of these systems in detail and provide an overiew on the physical implications of the results.

\section{Near-IR Adaptive Optics Observations}

We obtained adaptive optics (AO) images of multiple systems in the Taurus star forming region using the near-infrared camera NIRC2 \citep{wizinowich00} on the 10-m Keck II Telescope at the W.\ M.\ Keck Observatory.  The images were taken with the narrow-field camera that has a field of view of 10$\arcsec$.  During each observation, the target itself was used as the on-axis guide star.

\begin{figure}
	\scalebox{0.68}{\includegraphics{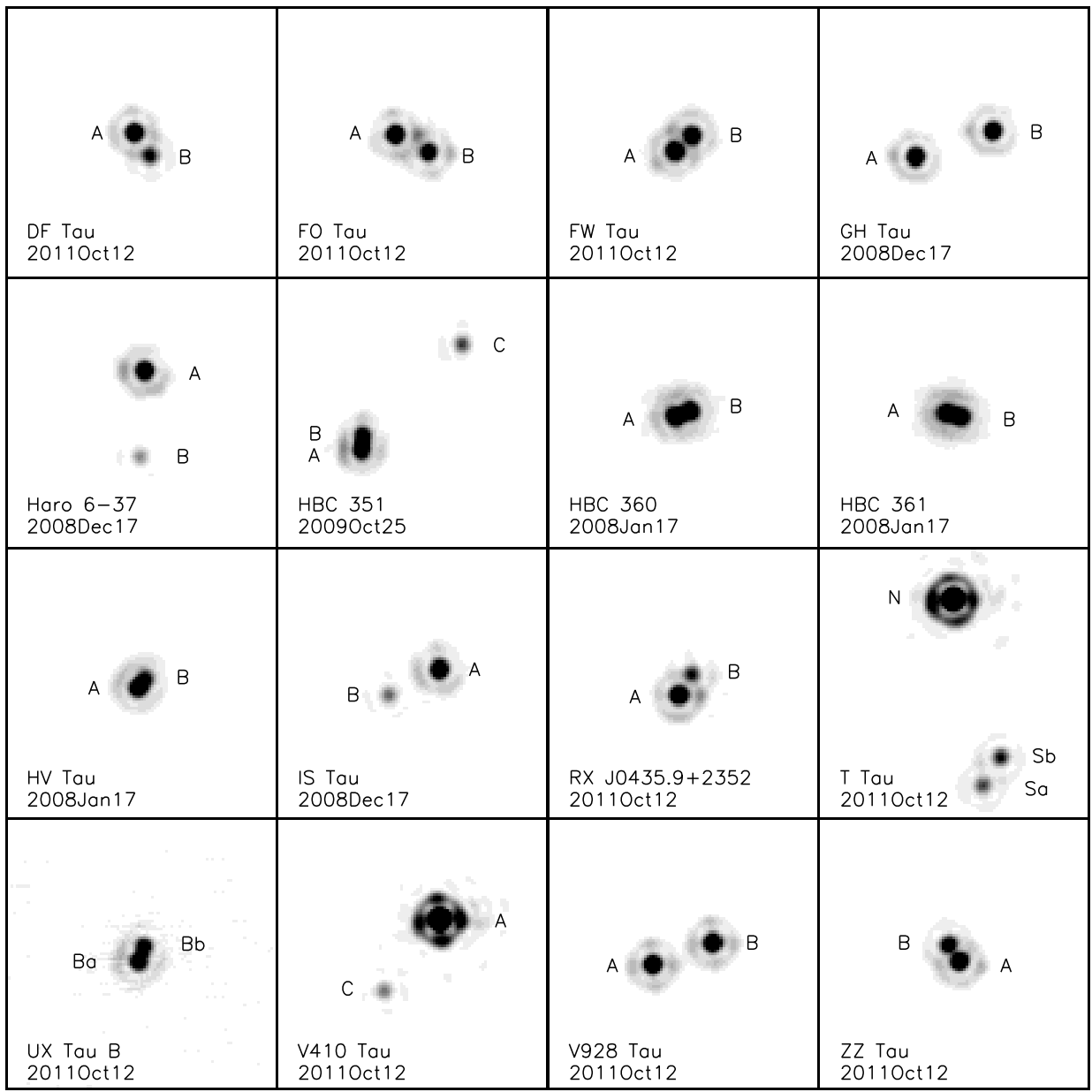}}
	\caption{Keck NIRC2 AO images of PMS binaries in the Taurus star forming region.  Each panel is $\sim 1 \arcsec$ wide.  North is up and east is to the left.  Fluxes are scaled by 0.4 times the maximum value, except for T Tau and V410 Tau which are scaled to 0.1 times the maximum.  Wide companions with separations larger than 1$''$ are not shown.}
\label{fig.images}
\end{figure}

Table~\ref{tab.aolog} presents a log of the AO observations we have taken since \citet{schaefer06}.  We list the UT date and time of observation, target name, filter, AO rate, integration time per exposure, and number of images taken in each filter. Each image was composed of 10 coadded exposures, except for the images on UT 2006 Dec 6 which used 5 coadds.  We obtained sets of dithered images with a 2$\arcsec$ offset using a 5-point dither pattern in 2006 through 2008 Jan and a 3-point dither pattern in 2008 Dec through 2014.  For wide systems, like UX Tau and HBC 360/361, we also made use of 2-point dither patterns to keep all components on the detector.
The images were flatfielded using dark-subtracted, medianed dome flats obtained on the nights of observation.  Pairs of dithered images were subtracted to remove the sky background.  Figure~\ref{fig.images} shows examples of co-added images for the PMS binaries in our sample with the relevant components are identified.  

For triple and higher order multiple systems, we used the wide ``single'' component as a simultaneous point spread function (PSF) to model the close pair.  For wider binaries where the airy rings of the PSF for each component did not overlap ($\gtrsim 200$ mas), we used the primary component as the PSF to measure the relative separation of the secondary.  For close binaries with overlapping PSFs and no wide component in the field of view, we observed a separate single star PSF reference (e.g., DN Tau, HBC 352) either immediately before or after the target observations.  Using the same AO frame rate between the target and PSF star minimizes differences in the shape of the corrected PSFs, thereby improving the quality of the results from the PSF fitting routine.  We followed the approach of matching the frame rates more systematically during later epochs.  In the last column of Table~\ref{tab.aolog} we indicate the PSF used to model each binary.

We used the relevant PSFs to measure the separation, position angle, and flux ratio of the binaries in our sample following the technique described in \citet{schaefer06}.  For the close pairs, we extracted subarrays from the images centered on the pair, with widths of $0\farcs1 - 0\farcs3$.  We used the PSF to construct models of the close pair by searching through a grid of separations and flux ratios and selecting the solution where the $\chi^2$ between the data and model reached a minimum.  We used the IDL INTERPOLATE procedure to shift the PSF by sub-pixel intervals using cubic convolution interpolation.  We used a slanting plane to model any excess light in the background (this step is typically negligible unless light from a wider component is present in the extraction box).  For the wider binaries, we used the same procedure, but fit for the relative position and flux ratio of the secondary as a ``single'' star using the primary as the PSF.  We corrected the measured positions for geometric distortions in the detector, used a plate scale of 9.952 $\pm$ 0.001 mas pixel$^{-1}$, and subtracted $0\fdg252 \pm 0\fdg009$ from the raw position angles to correct for the orientation of the camera relative to true north \citep{yelda10}.  We determined uncertainties in the positions and flux ratios by analyzing multiple images individually and computing the standard deviation.  

Table~\ref{tab.sepPA} lists the Besselian year, separation, position angle measured east of north, and flux ratio for each binary observed.  For multiple systems, we list the measurements for each pair of components separately.  In addition to observations listed in Table~\ref{tab.aolog}, we also provide revised values for the NIRC2 measurements of DF Tau, T Tau, and ZZ Tau previously published in \citet{schaefer06}.  We applied the geometric distortion correction determined by \citet{yelda10} to the previously measured positions.  For T Tau, we also computed the uncertainties for the close pair directly rather than propagating the uncertainties derived from the widely spaced separations of S$-$Na and S$-$Nb.

\section{Analysis of Orbital Motion}
\label{sect.orbit}

For each multiple system in our sample, we searched the literature for previously published measurements of their separations.  
In the following subsections, we plot the orbital motion for the close binaries based on our AO observations presented in Table~\ref{tab.sepPA} and the collection of measurements in the literature.  We can group the binaries into three categories based on their observed motion: (1) Binaries with enough orbital coverage to compute preliminary orbital solutions (DF Tau, T Tau S, ZZ Tau, and HBC 351). (2) Binaries showing curvature or acceleration in their relative motion, but where we can only place a lower limit on the orbital period (FO Tau, FW Tau, HBC 361, HV Tau, RX J0435.9+2352, UX Tau B, V410 Tau).  Preliminary solutions will be possible with more complete coverage of their orbits.  (3) Systems whose motion is indistinguishable from linear motion; continued observations will reveal whether these systems are bound (Haro 6-37, HBC 360, GH Tau, IS Tau, V928 Tau).  We describe the methods used to analyze the orbital motion in each of these categories and then provide detailed notes on specific systems.

Mapping the visual orbit of a binary system provides a measurement of the period ($P$), time of periastron passage ($T$), eccentricity ($e$), angular semi-major axis ($a$), inclination ($i$), position angle of the line of nodes ($\Omega$), and the angle between the node and periastron ($\omega$).  For binaries with sufficient orbital coverage, we solved for the seven orbital parameters using a Newton Raphson technique.  Uncertainties were computed from the diagonal elements of the covariance matrix.  The preliminary orbits for DF Tau, T Tau S, ZZ Tau, and HBC 351 are presented in Table~\ref{tab.orbpar}.  Coverage for three out of four of these orbits does not quite span 180$^\circ$ yet.  Therefore, the orbits presented here represent a reasonable description of their motion in the near future, although the orbital parameters may be subject to revision in the longer-term \citep[e.g.,][]{hartkopf01}.  In addition to computing formal orbital fits, we also used the approach outlined below to explore the $\chi^2$ parameter space and investigate the range of possible orbits that are consistent with the data.

\begin{figure}
	\scalebox{0.9}{\includegraphics{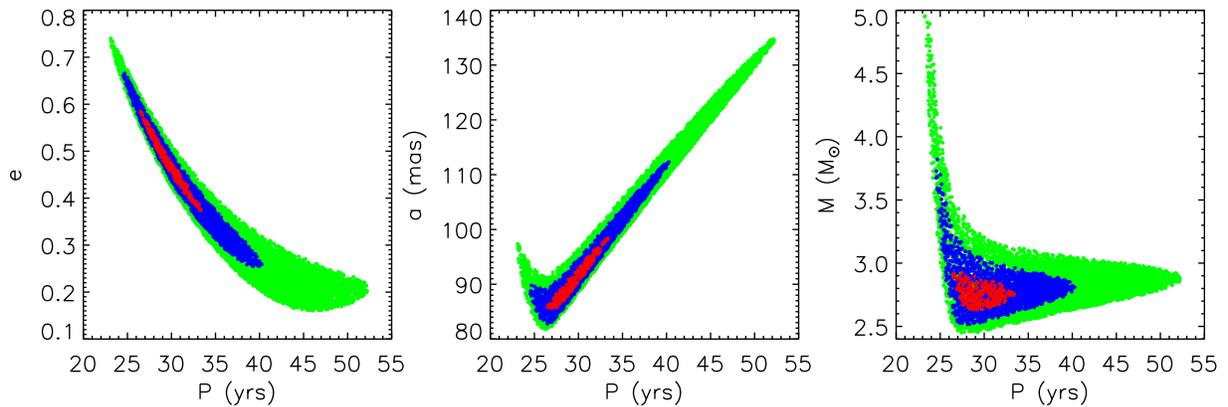}}
	\caption{Cross-cuts through the $\chi^2$ surface showing the $\Delta \chi^2 = 1$ (red), 4 (blue), and 9 (green) confidence intervals.  These plots were generated by searching for 10,000 orbital solutions for T Tau Sa-Sb by randomly varying $P, T$, and $e$ and optimizing the fit for the remaining parameters.  The projection of the $\Delta \chi^2 = 1$ surface onto each parameter corresponds to the 1\,$\sigma$ uncertainty when each parameter is considered individually.}
\label{fig.chi2}
\end{figure}

For binaries where only a small arc of the orbit is covered, we used the approach outlined by \citet{schaefer06} to randomly search the parameter space and explore the range of orbits that are consistent with the data.  We selected values of $P$, $T$, and $e$ at random from possible ranges that included periods out to 500 years, time of periastron that covered the full range of the chosen orbital period, and eccentricities from 0 to 0.99.  For each set of values for $P, T,$  and $e$, a least-squares fit to the data was performed to determine the optimal values for the Thiele-Innes elements ($A,B,F,G$) as outlined by \citet{hartkopf89} and \citet{mason99}.  The data were weighted by their respective measurement errors and the $\chi^2$ between the calculated $(x_c,y_c)$ and observed positions $(x_o,y_o)$ were computed from
\begin{equation}
\chi^2 = \sum \left(\frac{(x_o - x_c)^2}{\sigma_x^2} + \frac{(y_o - y_c)^2}{\sigma_y^2}\right) 
\end{equation}
where the errors, $\sigma_x$ and $\sigma_y$, were propagated from the measurement uncertainties listed in Table~\ref{tab.sepPA}.  After performing a grid search to determine the minimum $\chi^2$, we searched randomly for 10,000 possible solutions within $\Delta\chi^2 = 9$.  During this procedure, we scale the $\chi^2$ values so that the reduced $\chi^2_\nu = 1$ at the minimum.  The projection of the $\Delta\chi^2 = 1$ surface onto each of the orbital parameters corresponds to the 1\,$\sigma$ uncertainty intervals, assuming that the parameters are independent \citep[][, Ch.15]{press92}.  Results from the $\chi^2$ search for orbital solutions for T Tau Sa-Sb are shown in Figure~\ref{fig.chi2}.  The 1\,$\sigma$ intervals agree quite nicely with the uncertainties determined from the covariance matrix in the formal fit.  The strong correlation between $P$ and $a$ allows us to derive useful estimates for the total mass of the binary, $M_{\rm tot} \sim a^3/P^2$ \citep{eggen67,schaefer06,lucy13}.
For binaries where the orbital coverage is less complete, the $\chi^2$ surfaces are not as well-defined and the best we can do is place a lower limit on the orbital period.

The motion observed for five binaries does not yet show curvature.  For these systems, we computed a linear least-squares fit to model the motion in RA and Dec.  We found that the $\chi^2$ from the linear fit was indistinguishable from that obtained from a binary orbit fit.  Continued observations will reveal whether the companions are physically bound, possible ejections, or chance alignments.  Assuming that the systems are bound, we used the statistical $\chi^2$ search procedure described above to place a lower limit on their orbital periods.

For the four binaries with orbital solutions, we list the total mass of the components in the last column of Table~\ref{tab.orbpar}.  The uncertainties in $M_{\rm tot}$ were computed from the 1\,$\sigma$ confidence intervals determined during the $\chi^2$ search which takes into account the correlations between $a$ and $i$.  In computing the total masses of DF Tau and ZZ Tau, we assumed the systems are located at the average distance to the Taurus star forming region of 140 pc \citep{kenyon94}.  For T Tau, we used the trigonometric parallax of $6.90 \pm 0.09$ mas ($147.6 \pm 0.6$  pc) measured by \citet{loinard07} using the VLBA.  As discussed by \citet{torres09}, precise parallaxes of individual members of the Taurus star forming region range from 130 to 160 pc, so the 
actual location of a given star in our sample is likely to be different from the average value of 140 pc. In Figure~\ref{fig.taurus} we plot members of the Taurus star forming region and highlight stars with known distance measurements and binaries with preliminary orbits.  As the three dimensional structure becomes better known, associating stars with particular subgroups within the Taurus region might provide an improved distance estimate compared with using the fiducial benchmark of 140 pc.

\subsection{Comments on Individual Objects}
\label{sect.comments}

\subsubsection{Binaries with preliminary orbits}
\label{sect.prelim}

{\it DF Tau.} --  
DF Tau is a classical T Tauri star \citep{herbig88} that was first resolved as a binary during a lunar occultation survey \citep{chen90}.  The spectral types of DF Tau A and B are M2.0 and M2.5, respectively \citep{hartigan03}.  The orbital motion of the DF Tau binary has been mapped through speckle interferometry, imaging with the {\it Hubble Space Telescope (HST)} , interferometry with the Fine Guidance Sensors (FGS) on {\it HST}, and AO imaging \citep{chen90, ghez93, ghez95, thiebaut95, white01, balega02, schaefer03, balega04, schaefer06, shakhovskoj06, balega07}.  Preliminary orbits were computed by \citet{thiebaut95}, \citet{tamazian02}, and \citet{shakhovskoj06}, with periods ranging from 74 to 92 years.  However, as discussed by \citet{schaefer06} the reliability of these results is limited by the sparse orbital coverage. 

\begin{figure}
	\scalebox{1.0}{\includegraphics{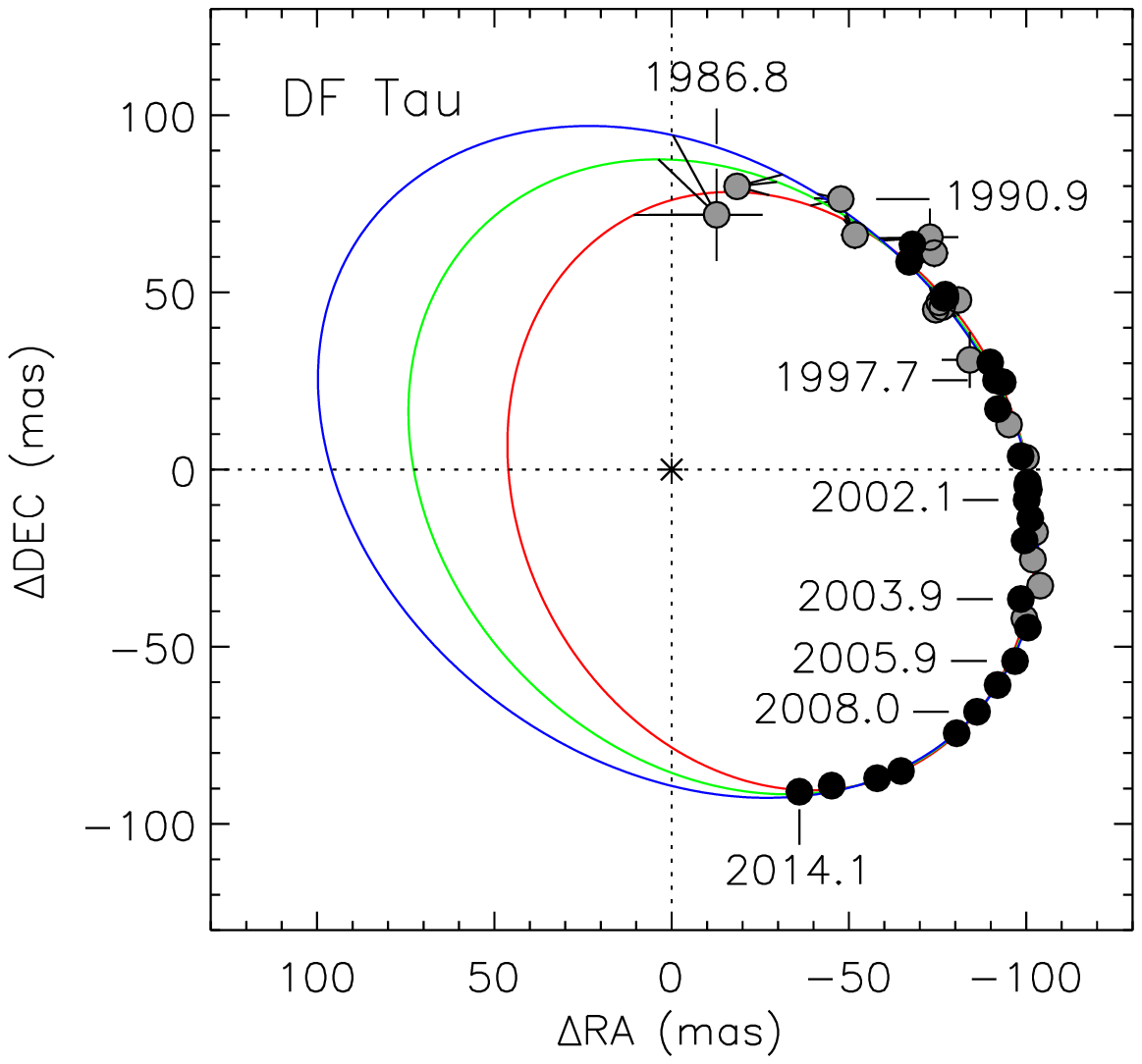}}
	\caption{Orbital motion measured for DF Tau.  The black circles are NIRC2 measurements presented in Table~\ref{tab.sepPA} and FGS measurements from \citet{simon96} and \citet{schaefer03,schaefer06}.  The grey circles are published values from the literature \citep{chen90, ghez95, thiebaut95, white01, balega02, balega04, shakhovskoj06, balega07}.  A formal fit yields an orbital period of 43.7 yr.  We plot examples of orbits at periods of 40 (red), 50 (green), and 60 yr (blue) that are consistent with the data within $\Delta \chi^2 = 9$ to demonstrate that longer orbital periods could be possible.  The first and last points provide the strongest constraints on limiting the range of possible orbital periods.}
\label{fig.dftau_orb}
\end{figure}

Figure~\ref{fig.dftau_orb} shows the orbital motion of DF Tau based on our AO observations and measurements in the literature (referenced above).  A formal orbital fit to DF Tau A-B gives an orbital period of 43.7 yrs and the parameters listed in Table~\ref{tab.orbpar}.  The 1\,$\sigma$ confidence intervals are well-defined, however we plot examples of longer period orbits in Figure~\ref{fig.dftau_orb} to demonstrate that we are only beginning to be able to determine a reliable orbit and that the parameters might need to be revised as more of the orbit is mapped.  The orbital solution is significantly different from previously published results because of the improved orbital coverage.  Assuming a distance of 140 pc, we derive $M_{\rm tot} = 1.17 \pm 0.13~M_\odot$.  

Interestingly, the best-fit orbital period closely matches the 44 yr timescale of the long-term photometric variability in DF Tau \citep[e.g.,][]{lamzin01}.  Using spatially resolved observations with the FGS, \citet{schaefer03} showed that the photometric variability can be attributed to the primary.  \citet{lamzin01} suggested that the long-term variability is caused by the circumstellar accretion rate onto the primary being modulated by the orbital motion of the companion.  The apparent agreement between the photometric and orbital periods is consistent with this interpretation.

{\it HBC 351.} --  
HBC 351 (HII 3197, TAP 9, V679 Tau, NTTS 034903+2431) was initially identified as a member of the Pleiades cluster based on its proper motion and position in the color-magnitude diagram \citep{hertzsprung47,johnson58}.  It also shows flare activity \citep{johnson58, jones81}.  Subsequently, it was identified as a candidate x-ray selected PMS star by \citet{feigelson87} and later classified as a naked T Tauri star by \citet{walter88} on the basis of Li I absorption and a radial velocity consistent with the Taurus-Auriga complex.  However, even though the x-ray luminosity and lithium abundance are on the high side, the values are consistent with other members of the Pleiades \citep{micela90, soderblom93}.  More recently, \citet{frink97} and \citet{luhman09} exclude it as a member of Taurus-Auriga on the basis of its proper motion being discrepant relative to the other nearby members of the group.  Moreover, the proper motion suggests that it has a high membership probability in the Pleiades \citep[e.g.,][]{schilbach95}.

HBC 351 was resolved as a 0\farcs6 binary by \citet{leinert93} and \citet{sterzik97}.  The system was discovered to be a triple by \citet{bouvier97}, with a close component at a separation of 110 mas.  We included HBC 351 in our program on binaries in the Taurus star forming region and present the results here, even though it is mostly likely a member of the Pleiades and not a PMS star.  We measured the separation of all three components during five epochs.  The results listed in Table~\ref{tab.sepPA} give the separation of the two close components relative to each other (similar brightness) and also the position of each close component relative to the fainter, wide tertiary component.  These wide separations are flipped by 180$^\circ$ relative to values in the literature which give the position of the faint component relative to the bright component. 

\begin{figure}
	\scalebox{1.0}{\includegraphics{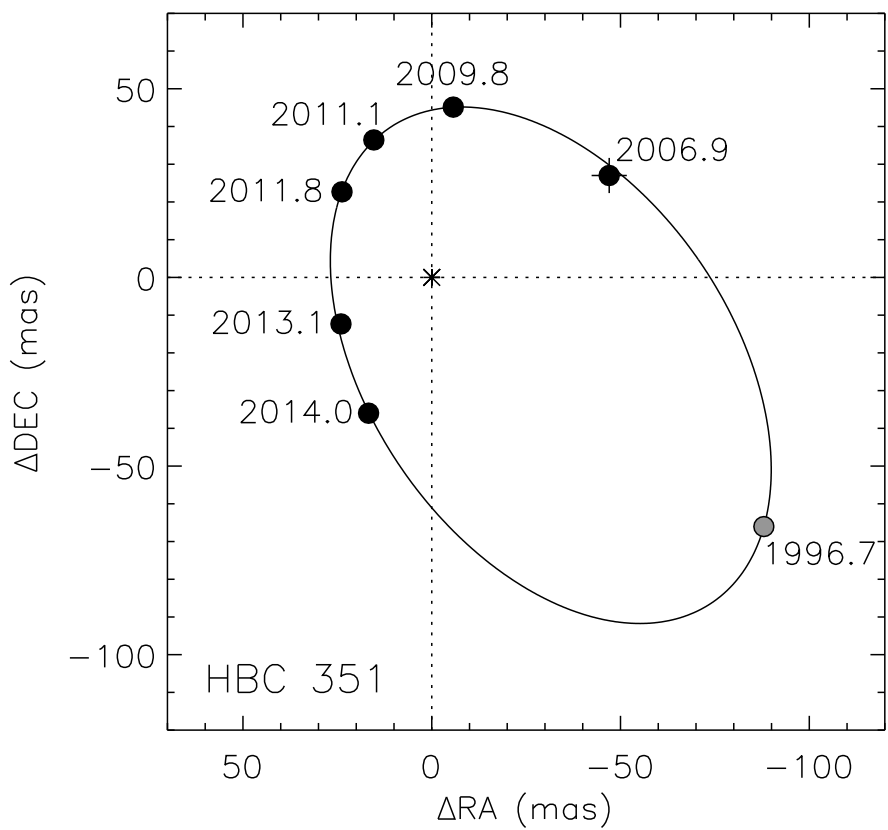}}
	\caption{HBC 351 is a triple system that is a probable member of the Pleiades (see text).  The black circles are NIRC2 measurements of the close pair, HBC 351 A-B, presented in Table~\ref{tab.sepPA}. The grey circle is the measurement from \citet{bouvier97}.  We computed a preliminary orbit with a period of 30.2 yr (solid line).  The fainter, wide tertiary component is located 0\farcs5 to the northwest and could be used to model the center of mass motion of the close pair as more of the orbit is mapped.}
\label{fig.hbc351_orb}
\end{figure}

In Figure~\ref{fig.hbc351_orb} we show the orbital motion of the close pair, HBC 351 A-B.  Our AO measurements combined with the results of \citet{bouvier97} allow us to compute a preliminary orbit with a period of 30.2 yrs (Table~\ref{tab.orbpar}).  With continued observations over a complete orbit, the mass ratio could be determined by modeling the center of mass motion of the close pair relative to the fainter tertiary.  If we adopt a parallax of 7.43 $\pm$ 0.17 mas for the Pleiades \citep[134.6 pc;][]{soderblom05}, we derive a total mass of $M_{\rm tot} = 1.42 \pm 0.10 M_\odot$ for HBC 351 A-B.  This is consistent with the sum of the individual masses of $M_A = 0.74 M_\odot$ and $M_B = 0.70 M_\odot$ determined by \citet{soderblom93} using the mass-luminosity relations in \citet{henry93}.  The revised  Hipparcos parallax for the Pleiades of 8.32 $\pm$ 0.13 mas \citep[120.2 pc;][]{vanleeuwen09}, gives a total mass of $M_{\rm tot} = 1.01 \pm 0.05 M_\odot$ which is not consistent with the corresponding photometric masses of the components.

{\it T Tau.} --  
T Tau is a young hierarchical triple system.  T Tau N is visible at optical wavelengths and has a spectral type of K0 \citep{basri90}.  The infrared companion, T Tau S, is located $\sim 0\farcs7$ to the south of T Tau N \citep{dyck82}.  T Tau S was found to have a close companion at a projected separation of $\sim 0\farcs05$ through speckle interferometry \citep{koresko00}.  T Tau Sa is likely an intermediate mass star while T Tau Sb has a spectral type of M1 \citep{duchene02,duchene05}. T Tau Sa and T Tau Sb remain optically invisible down to a limiting magnitude of $\sim$ 19.6 \citep{stapelfeldt98}.  Modeling of its spectral energy distribution suggests that T Tau S is the most luminous component with a large visual extinction from circumstellar material \citep{koresko97}.  Silicate and water-ice features indicate high-levels of foreground material toward T Tau Sa and Sb \citep{ghez91,beck04}.  Additionally, CO absorption in the spectrum of T Tau Sa suggests the presence of a circumstellar disk viewed nearly edge-on \citep{duchene05}.  

\citet{duchene06} and \citet{kohler08} present orbit fits that describe the motion of T Tau Sb relative to T Tau Sa.  By modeling the motion of each component relative to T Tau N, they also measure the motion of Sa and Sb relative to their center of mass.  Combining the orbital solution with the VLBA parallax of $6.90 \pm 0.09$ mas \citep[$147.6 \pm 0.6$  pc;][]{loinard07}, \citet{kohler08} derive individual masses of $M_{\rm Sa} = 2.1 \pm 0.2~M_\odot$ and $M_{\rm Sb} = 0.8 \pm 0.1~M_\odot$.  Both orbit fits confirm that T Tau Sa is the most massive component in the system.

\begin{figure}
	\scalebox{1.0}{\includegraphics{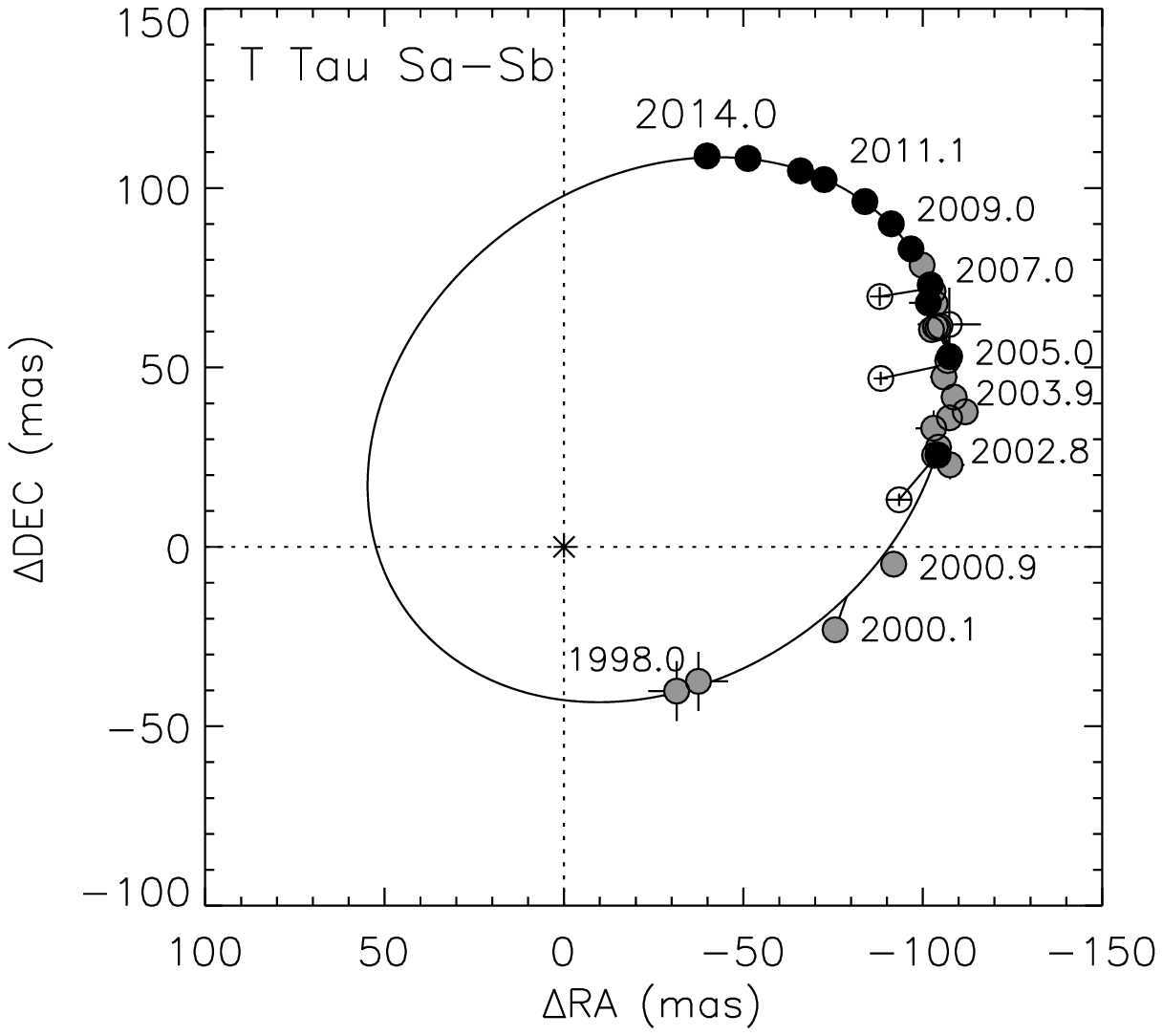}}
	\caption{Orbital motion measured for T Tau Sa-Sb.  The black circles are NIRC2 measurements presented in Table~\ref{tab.sepPA}.  The grey circles are published values from the literature \citep{kohler00, koresko00, duchene02, furlan03, beck04, duchene05, duchene06, mayama06, schaefer06, kohler08, skemer08, ratzka09}. Overplotted is the best fit orbit with a period of 29 yr.  Four measurements with large residuals \citep{mayama06, skemer08, ratzka09} were not included in the fit; these are plotted as open circles.  T Tau S is located 0\farcs7 south of the brightest optical/near-infrared component, T Tau N.}
\label{fig.ttau_orb}
\end{figure}

Figure~\ref{fig.ttau_orb} shows the orbital motion of T Tau Sb relative to Sa, based on our AO observations and published measurements in the literature \citep{kohler00, koresko00, duchene02, furlan03, beck04, duchene05, duchene06, mayama06, schaefer06, kohler08, skemer08, ratzka09}.  A formal fit to the orbit for T Tau Sa-Sb gives a period of 29 yr (see Table~\ref{tab.orbpar}).  The $\chi^2$ search results (Fig.~\ref{fig.chi2}) indicate that the $\Delta \chi^2 = 1$ confidence intervals are reasonbly well-defined, although the individual parameters will likely be revised as more of a complete orbit is mapped.  The shape of the best fit orbit is different from those presented by \citet{duchene06} and \citet{kohler08}; this is primarily a result of the improved coverage as the binary moved from apastron back to smaller separations in 2008$-$2014.  We have not re-computed a three-body solution that takes into account the motion of Sa and Sb relative to their center of mass because the fit will likely continue to be revised until more of a complete orbit is mapped.  Using the VLBA parallax, we derive $M_{\rm tot} = 2.70 \pm 0.22~M_\odot$.  This is consistent with the total mass computed by \citet{kohler08}.

\begin{figure}
	\scalebox{1.0}{\includegraphics{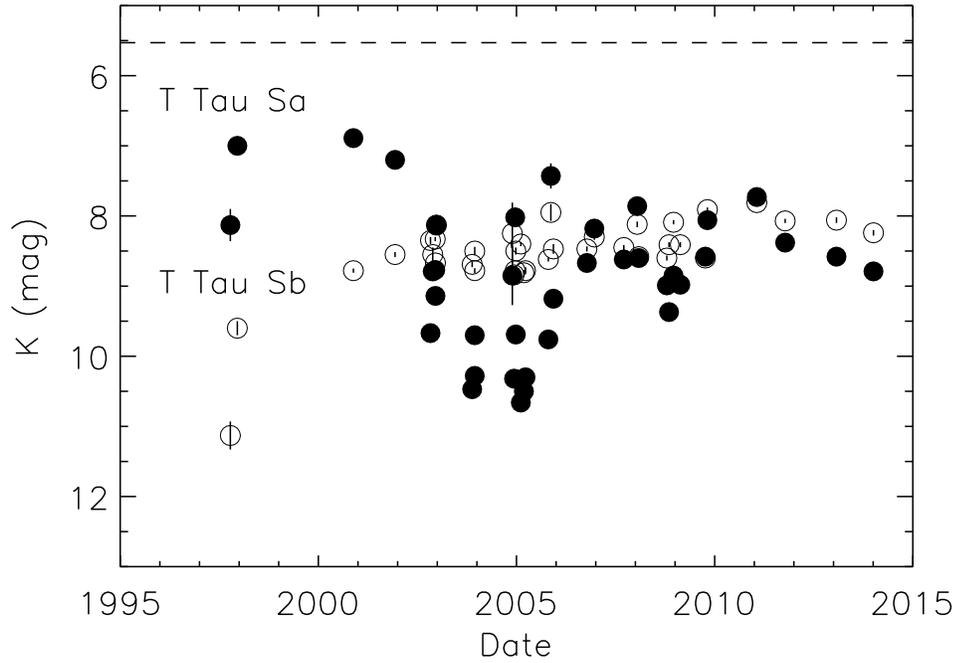}}
	\caption{$K$-band variability measured for T Tau Sa (filled circles) and T Tau Sb (open circles).  We derived the component magnitudes from the flux ratios and assume a constant magnitude of $K = 5.53 \pm 0.03$ mag for T Tau N \citep[dashed line;][]{beck04}.  We used the flux ratios from Table~\ref{tab.sepPA} and measurements in the literature \citep{koresko00,duchene02,duchene05,duchene06,furlan03,beck04,mayama06,schaefer06,vanboekel10}.}
\label{fig.ttaumag}
\end{figure}

\citet{beck04} determined that the flux of T Tau N remained constant from 1994 to 2002, with an average magnitude of $K = 5.53 \pm 0.03$ mag.  If we assume that the flux of T Tau N has remained the same over the last 10 years, then we can convert the relative fluxes of T Tau Sa and Sb into $K$-band magnitudes.  These results are plotted in Figure~\ref{fig.ttaumag}, along with magnitudes derived from the flux ratios presented in the literature \citep{koresko00,duchene02,duchene05,duchene06,furlan03,beck04,mayama06,schaefer06,vanboekel10}.  The large variability of T Tau Sa prior to 2006 was discussed by \citet{beck04} and \citet{duchene05}.  It appears that the variability of T Tau Sa has decreased in amplitude, however, this conclusion could be impacted by the less frequent sampling of the light curve in recent years. \citet{vanboekel10} noted the same trend and speculated that the previous period of high variability in Sa was caused by enhanced accretion in the disk triggered by the periastron passage of Sb.

\begin{figure}
	\scalebox{0.65}{\includegraphics{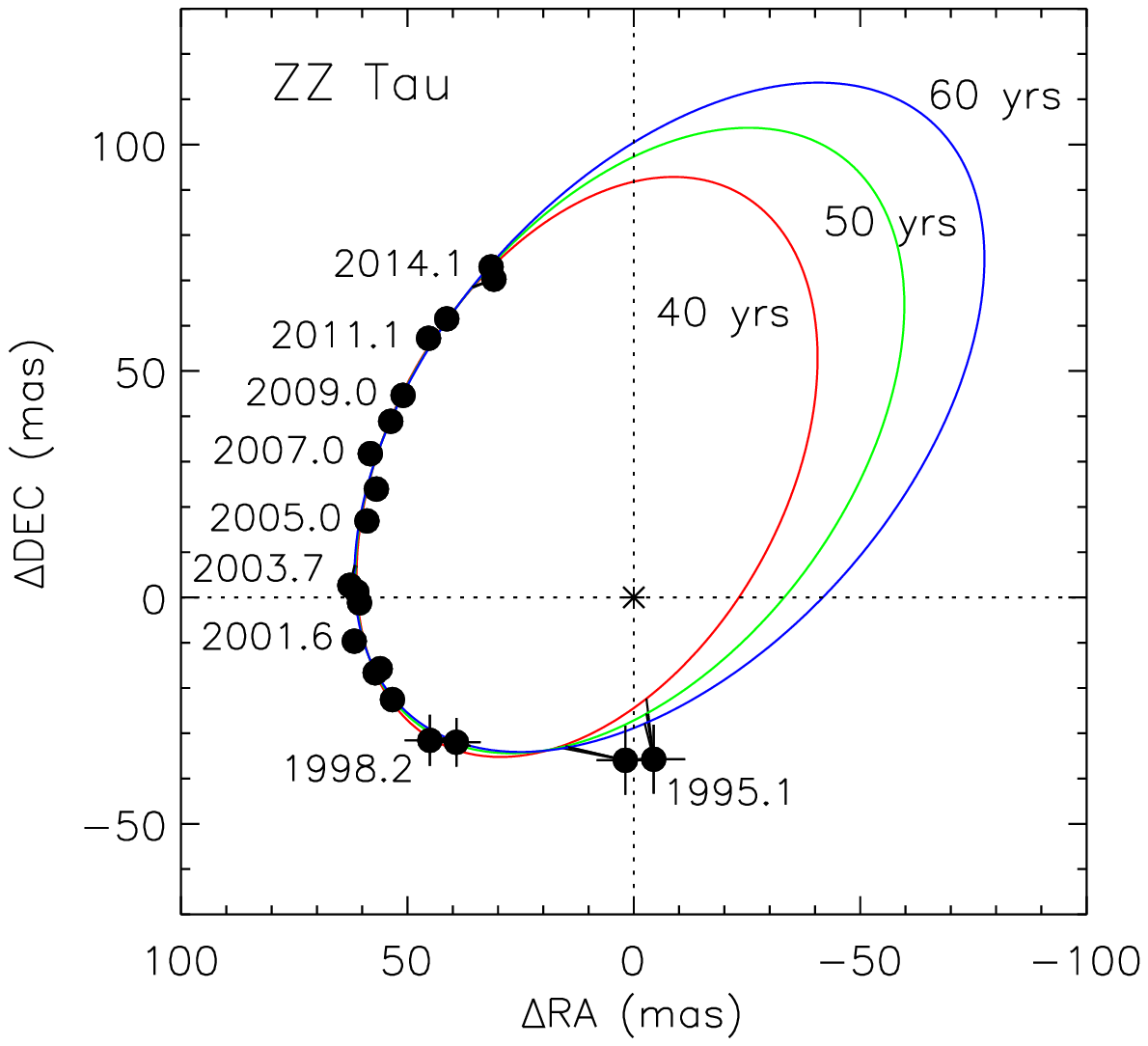}}
	\scalebox{0.65}{\includegraphics{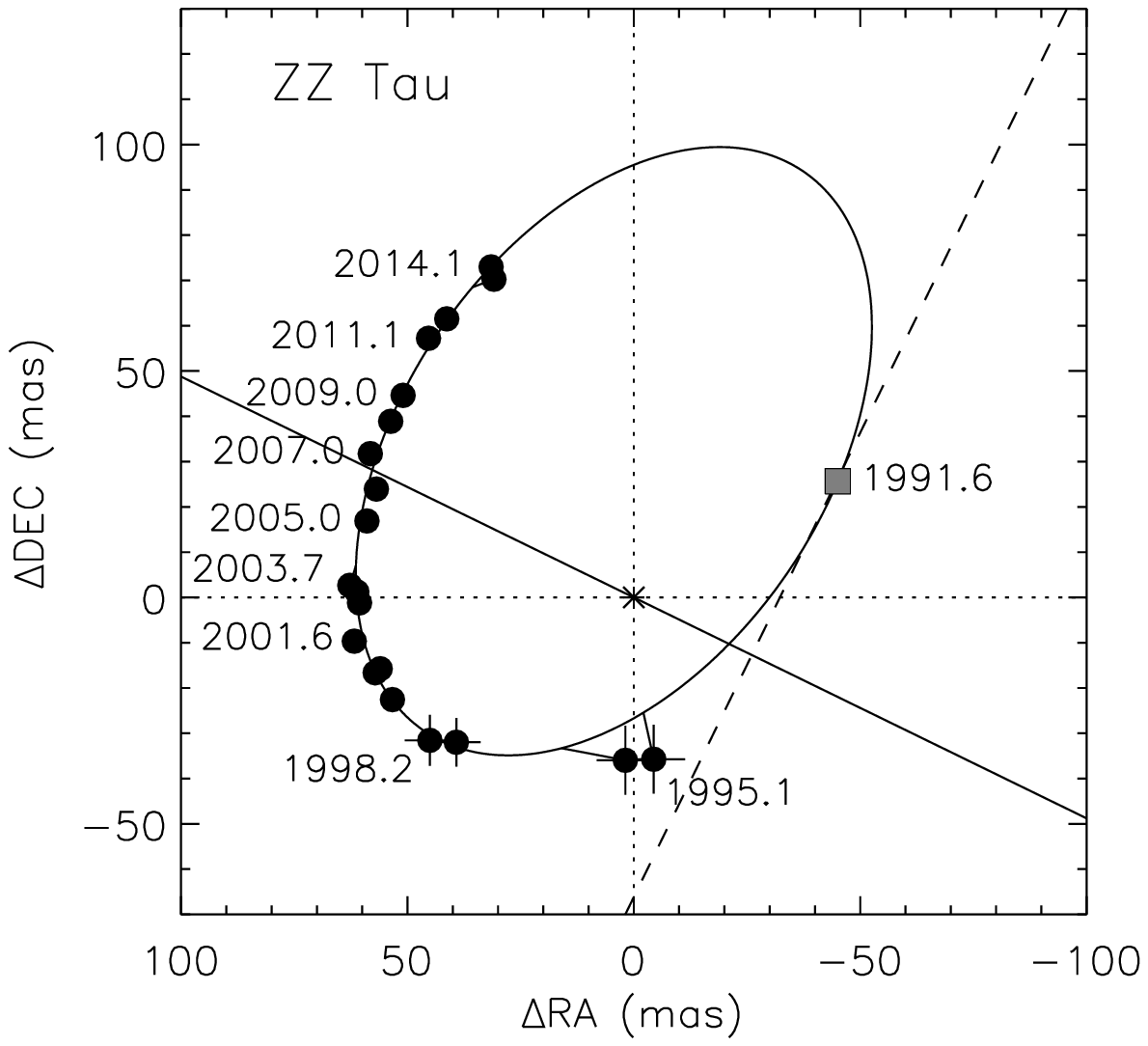}}
	\caption{Orbital motion measured for ZZ Tau.  The black circles are NIRC2 measurements presented in Table~\ref{tab.sepPA} and the FGS observations in \citet{schaefer03, schaefer06}.  Left: Examples of orbits at periods of 40 (red), 50 (green), and 60 yr (blue) that are consistent with the observed motion ($\Delta \chi^2 < 1.4$).  Right: Orbit including the lunar occultation measurement of ZZ Tau from 1991.6 \citep{simon95}. The solid line shows the direction of the lunar occultation along a position angle of 244$^\circ$. The dashed line shows the projected separation of 29 mas measured along this direction. The orbit must intersect this line to be consistent with the one-dimensional lunar occultation measurement.  Overplotted is the best-fit orbit (P $\sim$ 46.2 yr) based on the FGS, AO, and lunar occultation measurements. The position in the orbit at the time of the lunar occultation is marked by the shaded square.}
\label{fig.zztau_orb}
\end{figure}

{\it ZZ Tau.} --  
ZZ Tau is a classical T Tauri star with a spectral type of M3 \citep{herbig88}.  The system was discovered to be a binary through a lunar occultation \citep{simon95}.  Figure~\ref{fig.zztau_orb} shows the orbital motion measured for ZZ Tau, based on our recent and previously published AO and FGS measurements \citep{simon96, schaefer03, schaefer06}.  Fitting a formal orbit to these measurements yields an orbital period of $49 \pm 13$ yrs, with a 3\,$\sigma$ confidence interval from the $\chi^2$ search that extends out to 270 yrs.  

During the lunar occultation in 1991.6, \citet{simon95} measured a projected separation of 29 mas along a position angle of 244$^{\circ}$.  The orientation of the occultation is indicated in the right panel of Figure~\ref{fig.zztau_orb}.  We used a standard Newton Raphson approach to compute a simultaneous fit to the AO, FGS, and lunar occultation data.  Table~\ref{tab.orbpar} lists the best fit orbital parameters, with a period of $46.2 \pm 2.8$ yrs.  All of the parameters are consistent with the 1\,$\sigma$ uncertainties from the fit without including the lunar occultation.  However, the orbit differs significantly from our previously published solution which had a shorter period of $\sim 32$ yrs \citep{schaefer06}.  Even though the lunar occultation improves the time coverage, it provides only a one-dimensional constraint on the orbit.  Because of this and the fact that the companion continues moving toward larger separations, we suspect that the orbital parameters will be revised substantially in the future.  Nonetheless, the orbital parameters in Table~\ref{tab.orbpar} could be useful for computing the expected location of the binary over the next few years.  Assuming a distance of 140 pc, we derive $M_{\rm tot} = 0.83 \pm 0.16~M_\odot$ from the orbit fit to the AO, FGS, and lunar occultation data.  

\subsubsection{Binaries showing curvature}
\label{sect.curvature}

\begin{figure}
	\scalebox{1.0}{\includegraphics{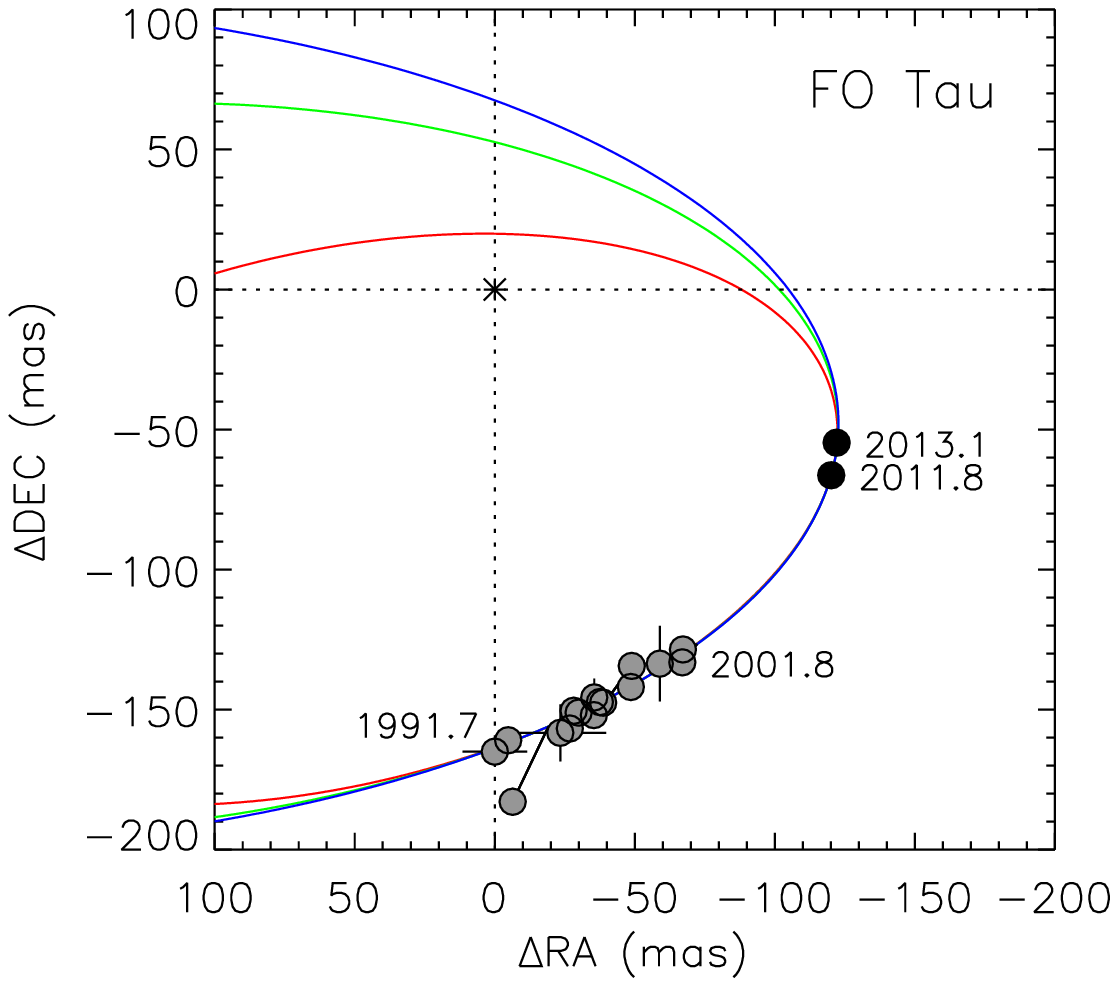}}
	\caption{Orbital motion measured for FO Tau.  The black circles are NIRC2 measurements presented in Table~\ref{tab.sepPA}.  The grey circles are published values from the literature \citep{leinert93, woitas01, ghez95, white01, hartigan03, tamazian02}.  Overplotted are three example orbits at periods of 100 (red), 200 (green), and 300 yr (blue) that are consistent with the observed motion.  A statistical analysis of orbital solutions that fit the data indicates that $P > 44$ yr.}
\label{fig.fotau_orb}
\end{figure}

{\it FO Tau.} --  
FO Tau is a binary containing two classical T Tauri stars with spectral types of M3.5 \citep{hartigan03}.  The measured separation ranges from 130 to 180 mas \citep[][; this work]{leinert93, woitas01, ghez93, ghez95, white01, hartigan03, tamazian02}.  Figure~\ref{fig.fotau_orb} shows the orbital motion for FO Tau based on our AO observations and previously published measurements.  The motion shows curvature and covers $\sim 66^\circ$ of the orbit with the addition of our AO measurements.  However, the coverage is not sufficient to determine a definitive orbital solution.  \citet{tamazian02} published a preliminary orbit for FO Tau with a period of 194 yr.  The specific parameters of the Tamazian et al.\ orbit are marginally consistent with our new measurements, but will require revision as more of the orbit is mapped.  In Figure~\ref{fig.fotau_orb} we overplot examples of possible orbits over a range of periods that are consistent with the data.  A statistical analysis of orbital solutions that fit the existing data indicates that the orbital period is greater than 44 yr, with the 1\,$\sigma$ confidence interval extending out to the search range of 500 yrs.

{\it FW Tau.} --  
FW Tau A-B is a close pair of two weak-lined T Tauri stars with spectral types of M5.5 \citep{white01}.  The measured separation of the close pair ranges from 70 to 160 mas.  \citet{white01} reported the detection of an additional faint companion, FW Tau C, at a separation of 2\farcs3 using {\it HST}.  In those observations, FW Tau C appeared brighter in the narrowband H$\alpha$ filter compared with the broadband filters.  The detection of the companion was confirmed by \citet{kraus14} using near-IR AO observations.  They speculated that FW Tau C (identified as FW Tau b in their paper) has a planetary mass based on its dereddened colors, although they do not rule out the possibility of a low mass star or brown dwarf surrounded by an edge-on disk.

\begin{figure}
	\scalebox{0.65}{\includegraphics{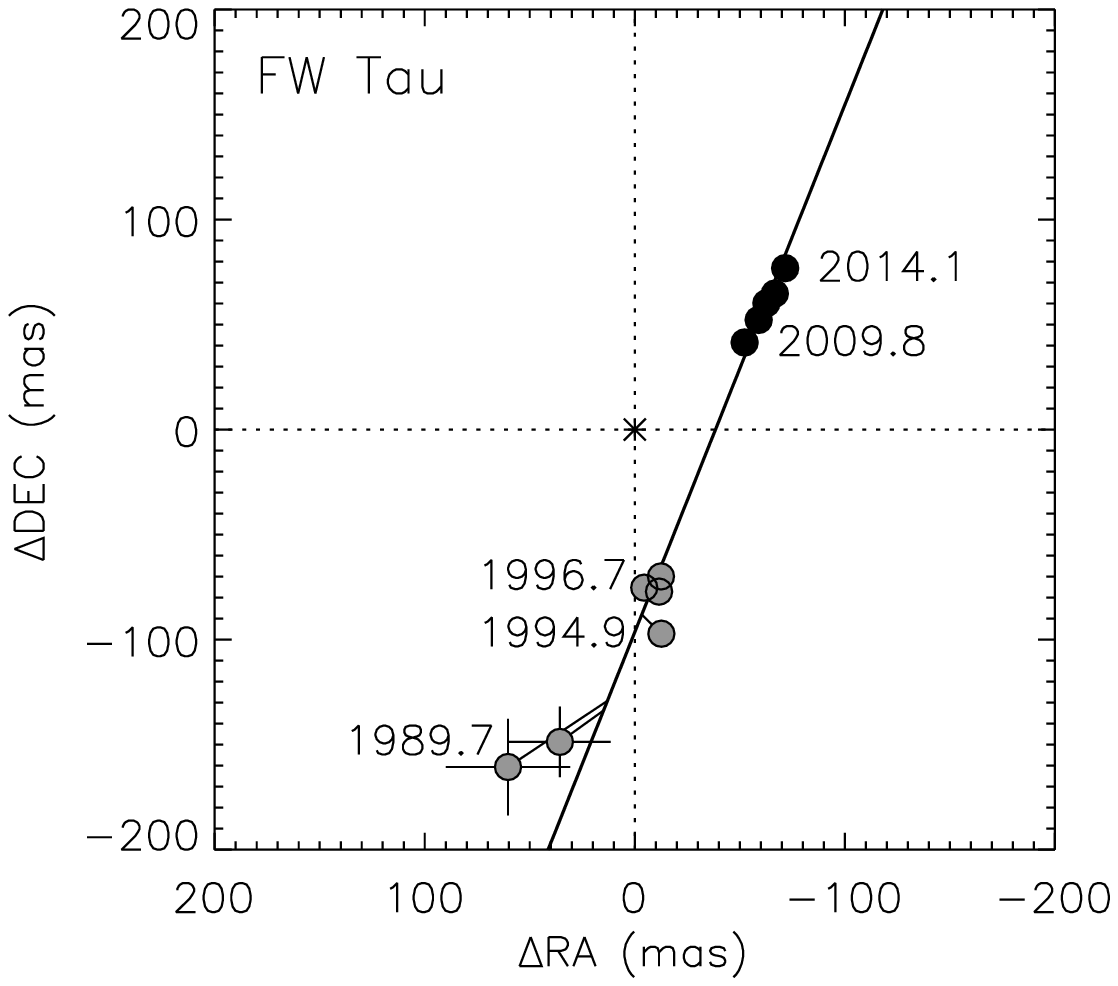}}
	\scalebox{0.65}{\includegraphics{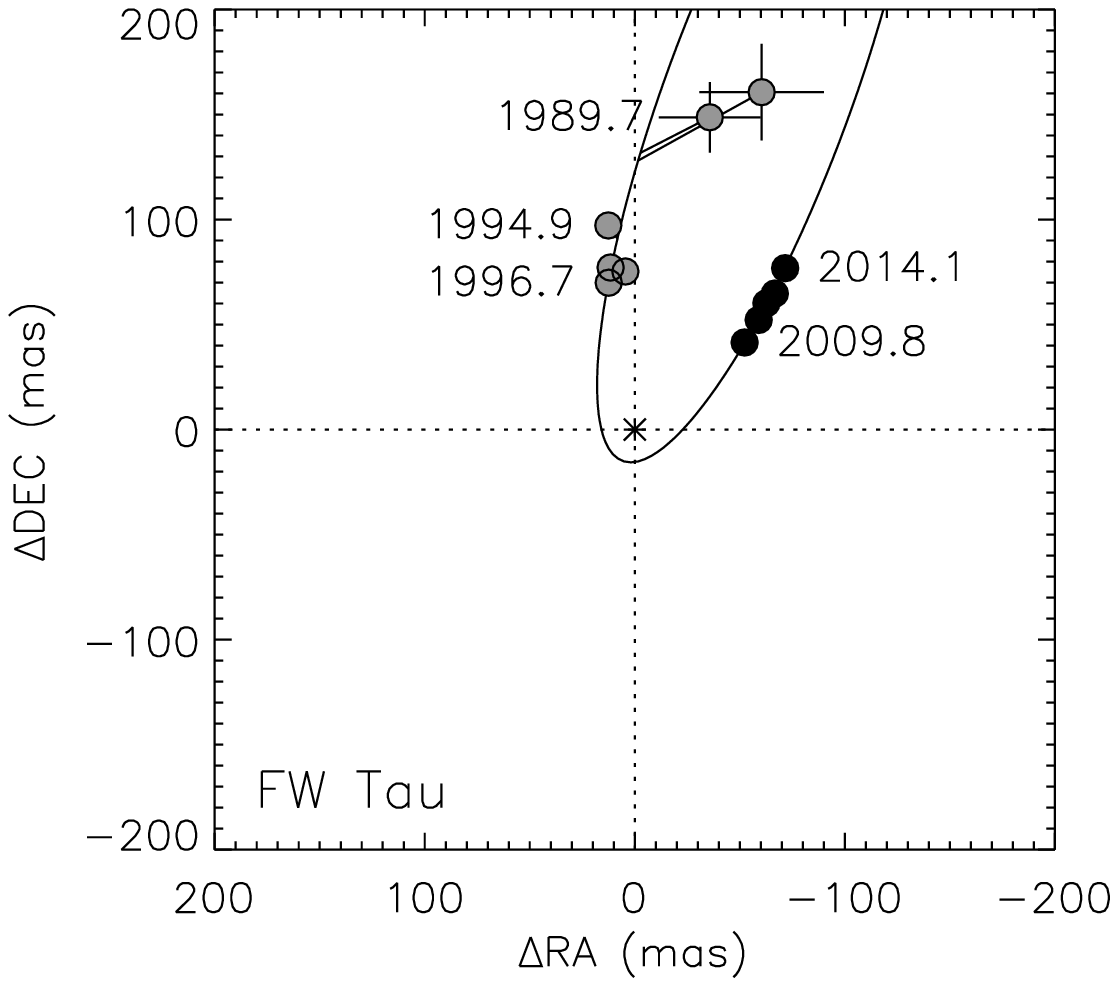}}
	\caption{Orbital motion measured for FW Tau. The black circles are NIRC2 measurements presented in Table~\ref{tab.sepPA}.  The grey circles are published values from the literature \citep{chen90, leinert91, simon92, woitas01, white01}.  The components are nearly equal in brightness, so there are two possible scenarios, depending on how we combine the earlier measurements with our more recent AO values.  The plot on the left shows the position angles as reported directly in the literature \citep[the position angle from][ has been flipped by 180$^\circ$ for consistency with the earlier measurements]{white01}.  This solution shows motion that is linear over the time frame of the observations.  The plot on the right shows the position angles from the 1990's flipped by 180$^\circ$.  This scenario produces orbital solutions that have lower $\chi^2$ values compared with the linear motion scenario on the left.  The example orbit on the right has a period of 200 yr.  More measurements are needed to determine which is correct.  In either scenario, a statistical analysis of orbital solutions that fit the data indicates that $P > 50$ yr.}
\label{fig.fwtau_orb}
\end{figure}

We detect FW Tau A-B in all of our AO images. On UT 2014 Jan 25, we took a series of deeper exposures to detect FW Tau C; the short exposures from earlier epochs were not sensitive enough to detect the faint companion.  The components of the A-B pair are nearly equal in brightness, with an average flux ratio of $0.96 \pm 0.02$ based on our AO images in the $K$-band.  This presents a possible ambiguity of 180$^\circ$ in position angle when matching our measurements with those published in the literature \citep{chen90, leinert91, simon92, woitas01, white01}.  Figure~\ref{fig.fwtau_orb} presents two scenarios to describe the orbital motion observed for FW Tau A-B.  The panel on the left shows motion that is effectively linear over the time-frame of the observations while the panel on the right displays an orbit where the position angle of the measurements prior to 1997 have been flipped by 180$^\circ$.  The orbit fit produces a smaller $\chi^2$ compared with the linear fit, but more measurements are needed to determine which is correct.  In either scenario, a statistical analysis of orbital solutions that fit the data indicates that the orbital period is longer than 50 yr.

{\it HBC 360/361.} --  
HBC 360 and 361 (NTTS 040142+2150W and NTTS 040142+2150E) are a pair of weak-lined T Tauri stars with spectral types of M3$-$M3.5 and are separated by 7\farcs2 \citep{herbig88,hartigan94,hartigan03}.  \citet{hartigan94} speculate that the stars are likely to be a physical pair based on their separation of $\sim$ 1200 AU compared with the typical projected separation for young stars in the Taurus star forming region.  On UT 2006 Dec 6, we discovered close companions to both HBC 360 and 361, with separations of 50 and 64 mas, respectively.  Therefore, HBC 360 and 361 form a quadruple system made up of two close pairs.

\begin{figure}
	\scalebox{1.0}{\includegraphics{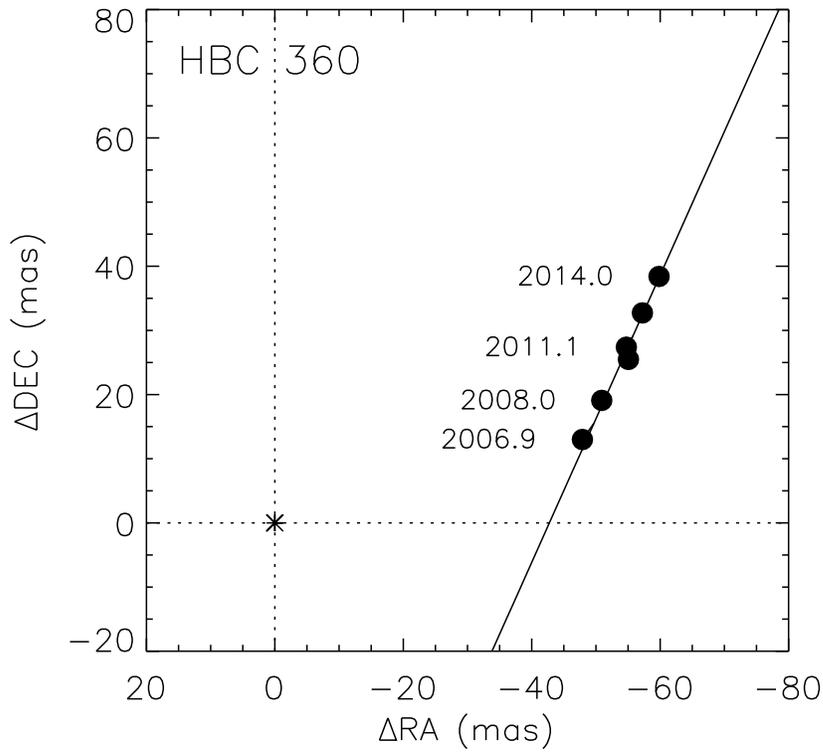}}
	\caption{Orbital motion measured for HBC 360.  The black circles are NIRC2 measurements presented in Table~\ref{tab.sepPA}.  The orbital motion is currently indistinguishable from linear motion (solid line).  If bound, then a statistical analysis of orbital solutions that fit the data indicates that $P > 25$ yr.}
\label{fig.hbc360_orb}
\end{figure}

Figure~\ref{fig.hbc360_orb} shows the orbital motion measured from our AO observations of the close pair in HBC 360.  We only have $\sim$ 7 yrs of coverage and thus far, the orbital motion for HBC 360 is indisguishable from linear motion.  If the pair is bound, then a statistical analysis of orbits that fit the data indicate that the orbital period must be greater than 25 yrs for HBC 360.

\begin{figure}
	\scalebox{1.0}{\includegraphics{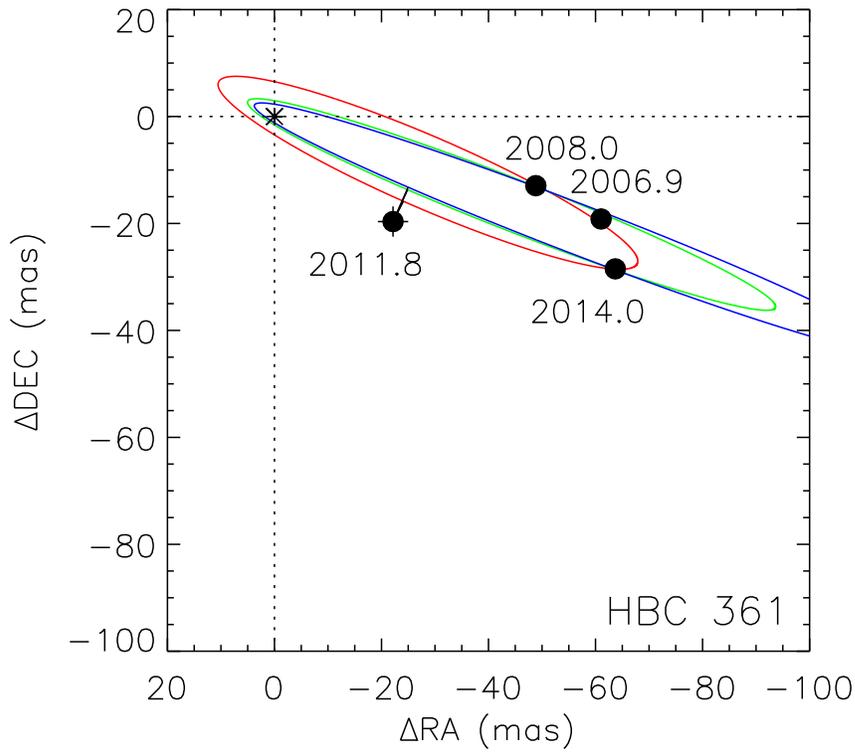}}
	\caption{Orbital motion measured for HBC 361.  The black circles are NIRC2 measurements presented in Table~\ref{tab.sepPA}.  Overplotted are three example orbits at periods of 10 (red), 20 (green), and 30 yr (blue) that are consistent with the observed motion.  A statistical analysis of orbital solutions that fit the data indicates that the 3\,$\sigma$ confidence interval for the orbital period extends beyond 500 yr.}
\label{fig.hbc361_orb}
\end{figure}

Figures~\ref{fig.hbc361_orb} shows the orbital motion measured from our AO observations of the close pair in HBC 361.  We resolved the pair during four of the six epochs.  On UT 2009 Oct 25 and 2011 Jan 24 we were unable to resolve the close components in HBC 361.  Visual inspection of the images did not reveal any elongation in the PSF of HBC 361 compared with the components in HBC 360 which were in the same field of view.  Fitting HBC 361 as a binary using a separate image of HBC 352 as a PSF reference, we estimate that the separation of HBC 361 was smaller than 20 mas on 2009.9 and smaller than 34 mas on 2011.1.  On UT 2011 Oct 12, the PSF of HBC 361 is visually elongated compared with the components in HBC 360 in the same field.  Fitting HBC 361 as a binary using a separate image of HBC 352 as a PSF, we measure a separation of $\sim$ 30 mas.  Separations on this order are at the limit of what we can resolve reliably without having a simultaneous PSF reference in the field of view.  However, given the fact that the position angle from the fit using the separate PSF matches the direction of elongation of HBC 361, we put a higher degree of confidence on these results compared with the results from the previous two years which showed no visual elongation.  The system was clearly resolved again on UT 2014 Jan 3.  Performing a $\chi^2$ search for orbital solutions, we found an eccentric, best-fit orbit with a period of $\sim$ 10 yr.  However, the 3\,$\sigma$ confidence interval extends beyond 500 yr.  Depending on the actual period, preliminary orbital parameters might be forthcoming with additional measurements in the near future.  

{\it HV Tau.} --  
HV Tau is a triple system.  The measured separation of the close pair, HV Tau A-B, ranges from 26 to 73 mas \citep[][; this work]{simon92, simon96, monin00, duchene10, kraus11}. \citet{white01} classified the A-B pair as a weak-lined T Tauri star with a spectral type of M2.  The third, fainter component, HV Tau C, lies at a separation of 4$''$ and a position angle of $\sim 45^\circ$ \citep{simon92, reipurth93, woitas98, magazzu94, monin00, stapelfeldt03, duchene10}; the position has stayed nearly constant from its discovery through the most recent observation.  HV Tau C is a low mass star (0.5$-$1.0 $M_\odot$) surrounding by an edge-on circumstellar disk \citep{monin00, stapelfeldt03, duchene10}.

We detected HV Tau A-B in our AO images, but did not detect any emission from HV Tau C.  The 4$''$ separation of C would have appeared on some of the dithered images, but the exposure times were optimized for the brighter A-B pair and not senstive enough to detect HV Tau C.  Using the standard deviation of the background to determine a 3\,$\sigma$ limit, we estimate that HV Tau C was at least $5.4$ mag fainter than HV Tau A in the $K$-band during our observations.  Previous measurements of the magnitude difference ranged from 3.3 to 5.2 mag at $K$ \citep{simon92, woitas98, monin00}.

\begin{figure}
	\scalebox{1.0}{\includegraphics{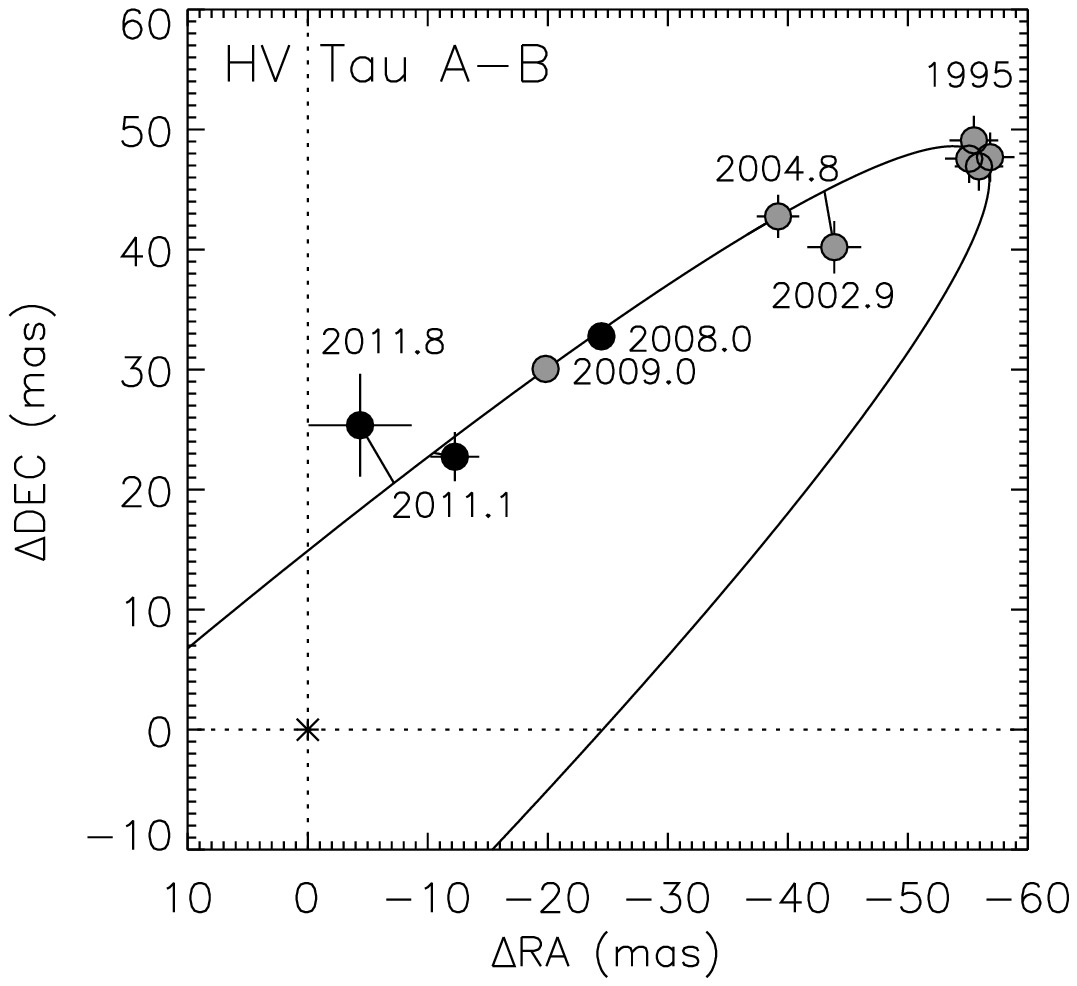}}
	\caption{Orbital motion measured for HV Tau A-B.  The black circles are NIRC2 measurements presented in Table~\ref{tab.sepPA}.  The grey circles are published values from the literature \citep{simon92, simon96, monin00, duchene10, kraus11}. Overplotted is an example orbit of 200 yr that is consistent with the data.  The orbit fit provides an improved $\chi^2$ compared with a linear fit, suggesting that we are measuring acceleration.  However, we do not have enough of the orbit mapped to determine reliable orbital parameters.  The position angles from of \citet{simon92} and \citet{simon96} have been flipped by 180$^\circ$.}
\label{fig.hvtau_orb}
\end{figure}

We show the orbital motion of HV Tau A-B based on our AO measurements and previously published results in Figure~\ref{fig.hvtau_orb}.  At first glance, the motion appears linear, however, the change in position over time suggests the presence of an acceleration term.  An orbital fit to the data produces a significantly lower $\chi^2$ compared with a fit to linear motion.  However, we do not have enough of the orbit mapped to determine reliable orbital parameters.  A statistical analysis of orbital solutions that fit the data indicates that the orbital period is longer than 27 yr.

\begin{figure}
	\scalebox{1.0}{\includegraphics{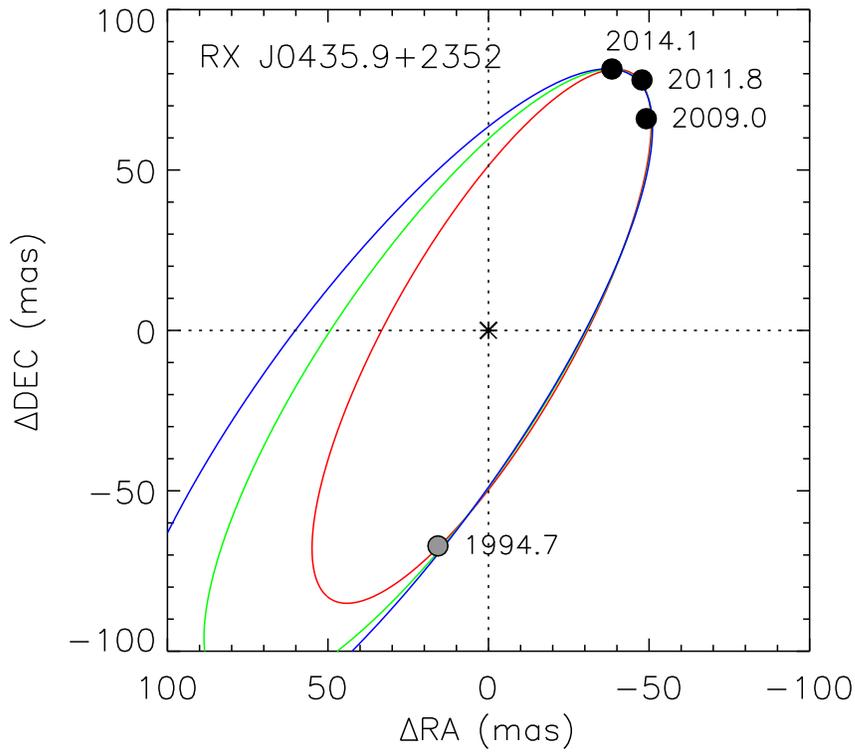}}
	\caption{Orbital motion measured for RX J0435.9+2352.  The black circles are NIRC2 measurements presented in Table~\ref{tab.sepPA}.  The grey circle is the measurement from \citet{kohler98}.  Overplotted are three example orbits at periods of 50 (red), 75 (green), and 100 yr (blue) that are consistent with the observed motion.  A statistical analysis of orbital solutions that fit the data indicates that $P > 22$ yr.}
\label{fig.rxj0435_orb}
\end{figure}

{\it RX J0435.9+2352.} --  
\citet{wichmann96} identified RX J0435.9+2352 as a weak-lined T Tauri star with a spectral type of M1.5 based on its x-ray emission and lithium abundance.  It was discovered to be a close binary with a separation of 70 mas by \citet{kohler98}.  We obtained three AO observations of RX J0435.9+2352 in 2008$-$2014 and found that the binary moved $\sim 168^\circ$ in position angle since its discovery.  Figure~\ref{fig.rxj0435_orb} shows the orbital motion for RX J0435.9+2352.  The binary motion shows curvature, but we do not have sufficient coverage to determine reliable orbital parameters.  We overplot a range of possible orbits that are consistent with the observed motion.  A statistical analysis of orbital solutions that fit the data indicates that the orbital period is longer than 22 yr.  

{\it UX Tau.} --  
UX Tau is a quaduple system.  UX Tau A is a classical T Tauri star with a spectral type of K5 \citep{hartigan94,white01}.  UX Tau B is a weak-lined T Tauri star with an M2 spectral type located 5\farcs9  west of UX Tau A.  UX Tau C is also a weak-lined T Tauri star with a spectral type of M5 located 2\farcs7 south of UX Tau A.  UX Tau B itself is a close pair (Ba-Bb) with separations ranging 30 to 140 mas \citep[][; this work]{duchene99, correia06}.  All four components of the system were detected in our AO images.

\begin{figure}
	\scalebox{1.0}{\includegraphics{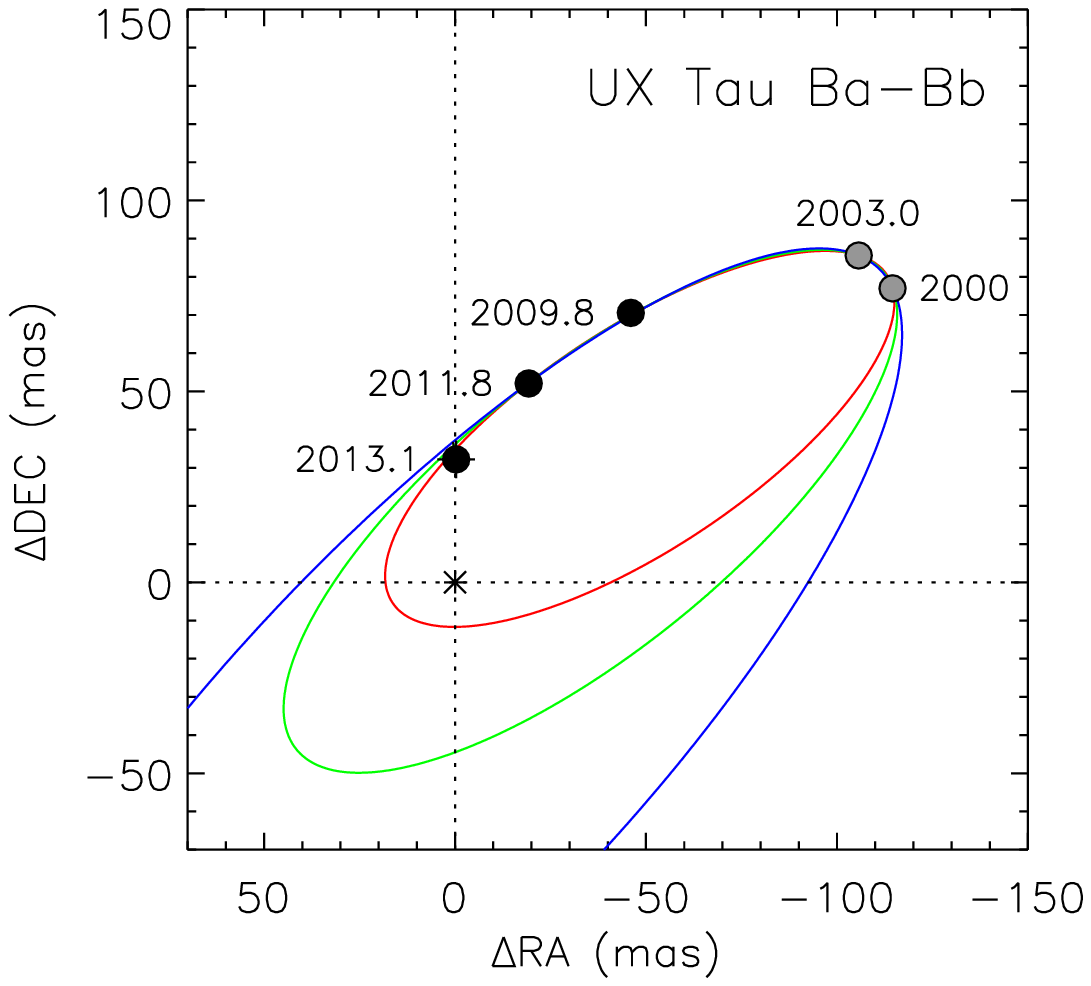}}
	\caption{Orbital motion measured for UX Tau Ba-Bb.  The black circles are NIRC2 measurements presented in Table~\ref{tab.sepPA}.  The grey circles are published values from the literature \citep{duchene99, correia06}. Overplotted are three example orbits at periods of 25 (red), 40 (green), and 100 yr (blue) that are consistent with the observed motion.  A statistical analysis of orbital solutions that fit the data indicates that $P > 20$ yr.  UX Tau B is located 5\farcs9 to the west of UX Tau A.  UX Tau C is located 2\farcs7 south of UX Tau A.}
\label{fig.uxtau_orb}
\end{figure}

Figure~\ref{fig.uxtau_orb} shows the orbital motion observed for UX Tau Ba-Bb based on our AO measurements and previously published values.  The coverage is nearly, but not quite good enough to compute a formal orbit fit.  The range of orbits plotted in Figure~\ref{fig.uxtau_orb} indicates that a reliable solution could be possible with continued observations over the next few years.  A statistical analysis of orbital solutions that fit the data indicates that the period is likely between 20$-$250 yr based on the 3\,$\sigma$ confidence intervals.

{\it V410 Tau.} --  
V410 Tau is a triple system.  The close pair, V410 Tau A-B, was discovered through near-infrared speckle observations \citep{ghez93}.  The close pair was initially detected at a separation of 130 mas in 1991 and at a smaller separation of 74 mas in 1994 \citep{ghez95}.  \citet{ghez97} and \citet{white01} reported the detection of a faint, third component (V410 Tau C) at a separation of 287 mas using {\it HST}, although the closer component was unresolved in those observations.  V410 Tau A is classified as a weak-lined T Tauri star with a spectral type of K4, while the fainter third component, V410 Tau C, has a spectral type of M5.5 \citep{white01}.  

\begin{figure}
	\scalebox{1.0}{\includegraphics{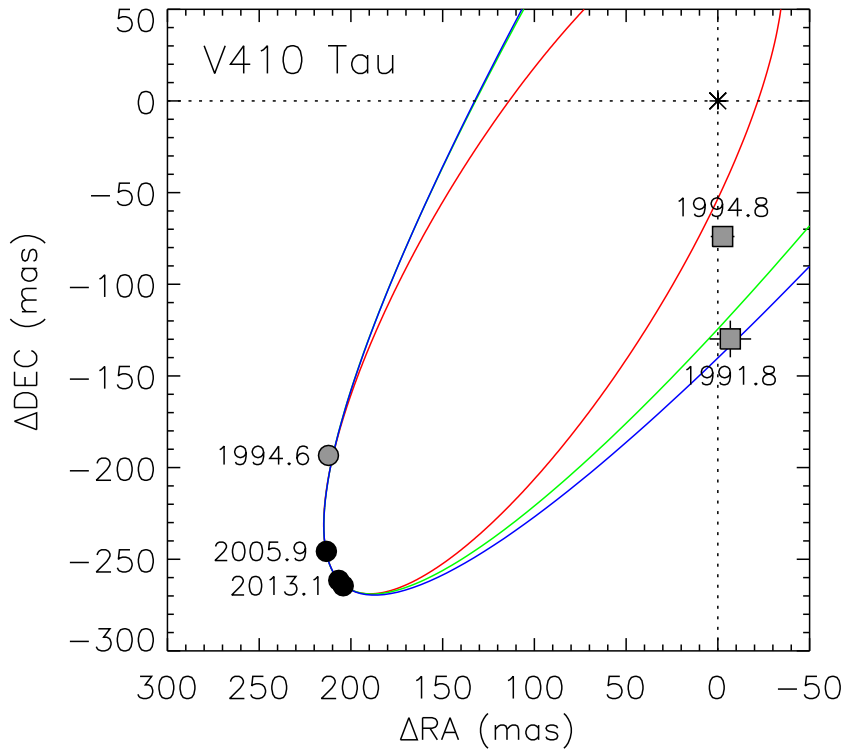}}
	\caption{Orbital motion measured for V410 Tau.  The black circles are NIRC2 measurements of V410 A-C presented in Table~\ref{tab.sepPA}.  The grey circle is from \citet{white01}.  Overplotted are three example orbits at periods of 100 (red), 200 (green), and 300 yr (blue) that are consistent with the observed motion. A statistical analysis of orbital solutions that fit the data indicates that $P > 68$ yrs.  The grey squares represent the speckle measurements of companion B detected by \citet{ghez95}.  We did not detect V410 Tau B in our AO images.}
\label{fig.v410tau_orb}
\end{figure}

We only detect V410 Tau A-C in our AO images.  According to \citet{ghez97}, the $K$-band magnitude difference between A-B is $\Delta K = 2.5$ mag, while that of A-C is $\Delta K = 3.0$ mag.  Therefore, all three components should be within the sensitivity limits of our images.  Our non-detection of A-B during the three epochs could indicate that the companion is much fainter than the previous detections, the separation is below the Keck-AO resolution, or it is blended into the diffraction ring.  If we assume that the B component continued on a linear path from its motion in 1991 to 1994 (see squares in Fig.~\ref{fig.v410tau_orb}), we expect that the component would be located 130 to 270 mas north of V410 Tau A in 2006$-$2013 and would still be located within the field of view of the AO images.

In Figure~\ref{fig.v410tau_orb} we also plot the motion of V410 Tau A-C based on our AO measurements and previously published values \citep{white01} as circles.  The motion shows curvature, but the coverage is too limited to warrant a full orbital solution.  We overplot a range of possible orbits that are consistent with the observed motion in Figure~\ref{fig.v410tau_orb}.  A statistical analysis of orbital solutions that fit the data indicates that the orbital period is longer than 68 yrs.

\subsubsection{Binaries not yet showing curvature}

\begin{figure}
	\scalebox{1.0}{\includegraphics{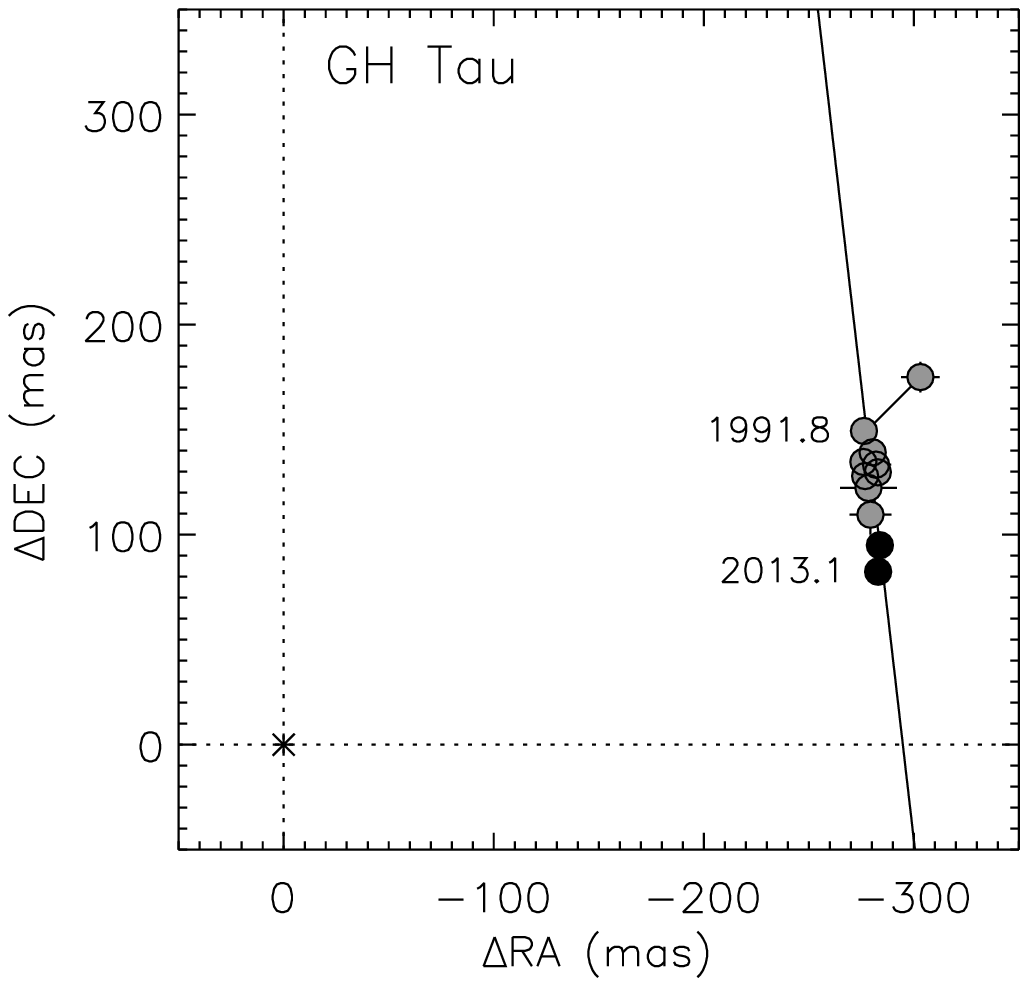}}
	\caption{Orbital motion measured for GH Tau.  The black circles are NIRC2 measurements presented in Table~\ref{tab.sepPA}.  The grey circles are published values from the literature \citep{leinert93, ghez95, white01, woitas01, hartigan03}.  The solid line shows a fit for linear motion.  If bound, then a statistical analysis of orbital solutions that fit the data indicates that $P > 95$ yr.}
\label{fig.ghtau_orb}
\end{figure}

{\it GH Tau.} --  
GH Tau is a binary containing two classical T Tauri stars with a spectral type of M2 \citep{hartigan03}.  The measured separation ranges from 330 to 350 mas \citep[][; this work]{leinert93, ghez93, ghez95, white01, woitas01, hartigan03}. It forms a wide multiple system with the triple V807 Tau that is located 22$''$ to the north \citep{hartigan94,kraus07}.  V807 Tau is outside the field of view of our NIRC2 images, however, we observed the triple separately and presented a complete orbit for the close pair in that system in \citet{schaefer12}.  Figure~\ref{fig.ghtau_orb} shows the orbital motion for GH Tau based on our AO observations and previously published measurements.  Fitting an orbit to the system provides a slightly improved $\chi^2$ compared with linear motion, but the coverage is so limited that we cannot place strong constraints on the orbital parameters.  A statistical analysis of orbits that fit the data indicates that the orbital period must be greater than 95 yr.

\begin{figure}
	\scalebox{1.0}{\includegraphics{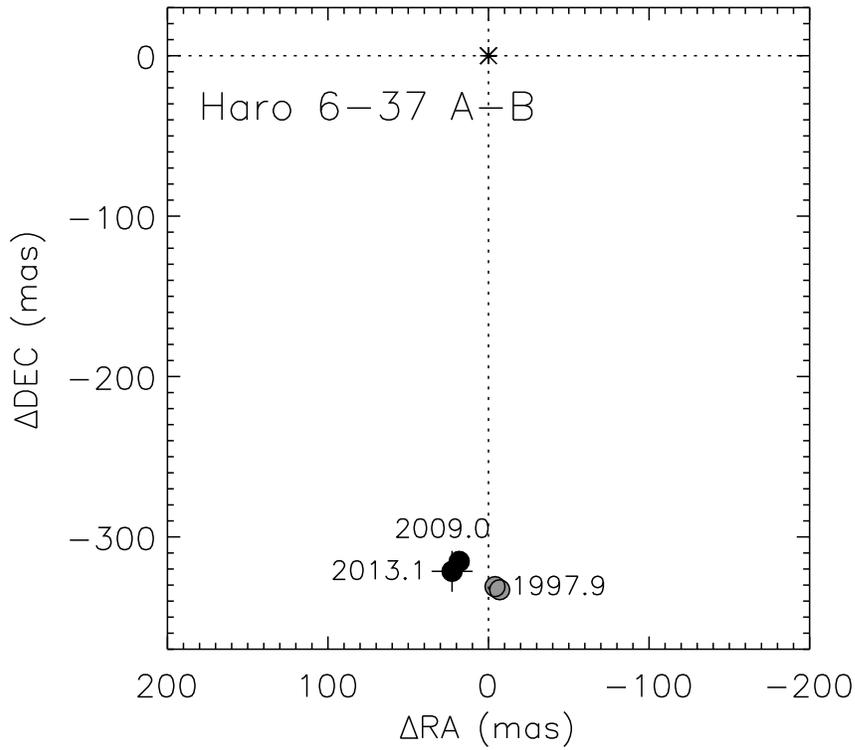}}
	\caption{Orbital motion measured for Haro 6-37 A-B.  The black circles are NIRC2 measurements of Haro 6-37 A-B presented in Table~\ref{tab.sepPA}.  The grey circles are published values from the literature \citep{duchene99, richichi99}.  The close pair is moving slowly relative to each other, so we cannot yet place limits on the orbital motion.  The faint tertiary, Haro 6-37 C, is located 2\farcs7 to the northeast of A.}
\label{fig.haro637_orb}
\end{figure}

{\it Haro 6-37.} --  
Haro 6-37 is a triple system.  The components of the wide pair, Haro 6-37 A-C, are separated by 2.6$''$ and a position angle of $\sim 38^\circ$ \citep[e.g.,][]{moneti91, reipurth93}.  Haro 6-37 A and C are identified as classical T Tauri stars with spectral types of K7 and M1, respectively \citep{hartigan94, white01}.  The close companion, Haro 6-37 B, is separated from A by $\sim$ 300 mas \citep{duchene99, richichi99}.  We detect all three components in our AO images.  Figure~\ref{fig.haro637_orb} shows the orbital motion for Haro 6-37 A-B based on our AO observations and previously published measurements.  The close pair is moving very slowly relative to each other, so we cannot yet place limits on their orbital motion.

\begin{figure}
	\scalebox{1.0}{\includegraphics{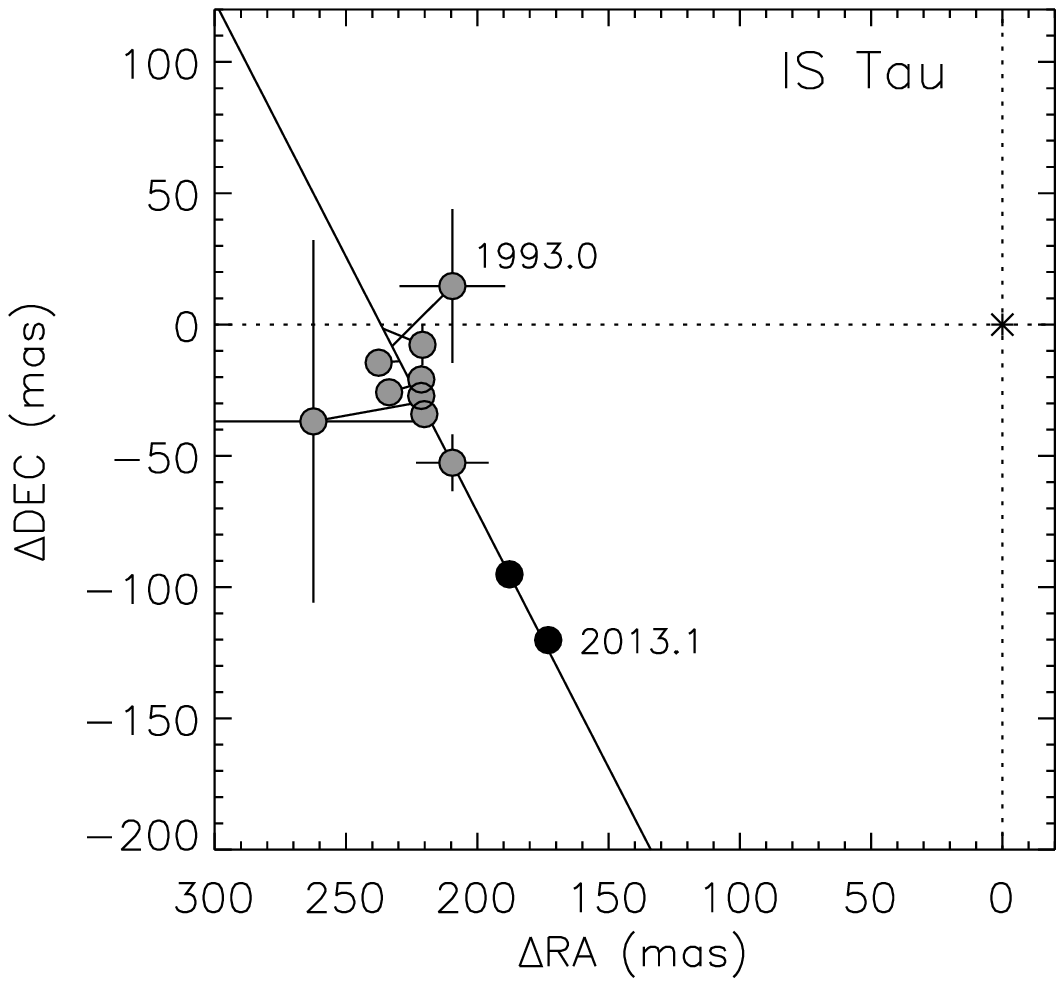}}
	\caption{Orbital motion measured for IS Tau.  The black circles are NIRC2 measurements presented in Table~\ref{tab.sepPA}.  The grey circles are published values from the literature \citep{leinert93, ghez93, white01, woitas01, hartigan03}.  The solid line shows a fit for linear motion.  If bound, then a statistical analysis of orbital solutions that fit the data indicates that $P > 57$ yr.}
\label{fig.istau_orb}
\end{figure}

{\it IS Tau.} --  
IS Tau is a binary system.  IS Tau A is a classical T Tauri star with a spectral type of M0 and IS Tau B is a weak-lined T Tauri star with a spectral type of M3.5 \citep{hartigan03}.  The measured separation of the binary ranges from 210 to 240 mas \citep[][; this work]{leinert93, ghez93, white01, woitas01, hartigan03}.   Figure~\ref{fig.istau_orb} shows the orbital motion for IS Tau based on our AO observations and previously published measurements.  Currently, the orbital motion is indistinguishable from linear motion.  A statistical analysis of orbits that fit the data indicates that the orbital period is longer than 57 yr.

\begin{figure}
	\scalebox{1.0}{\includegraphics{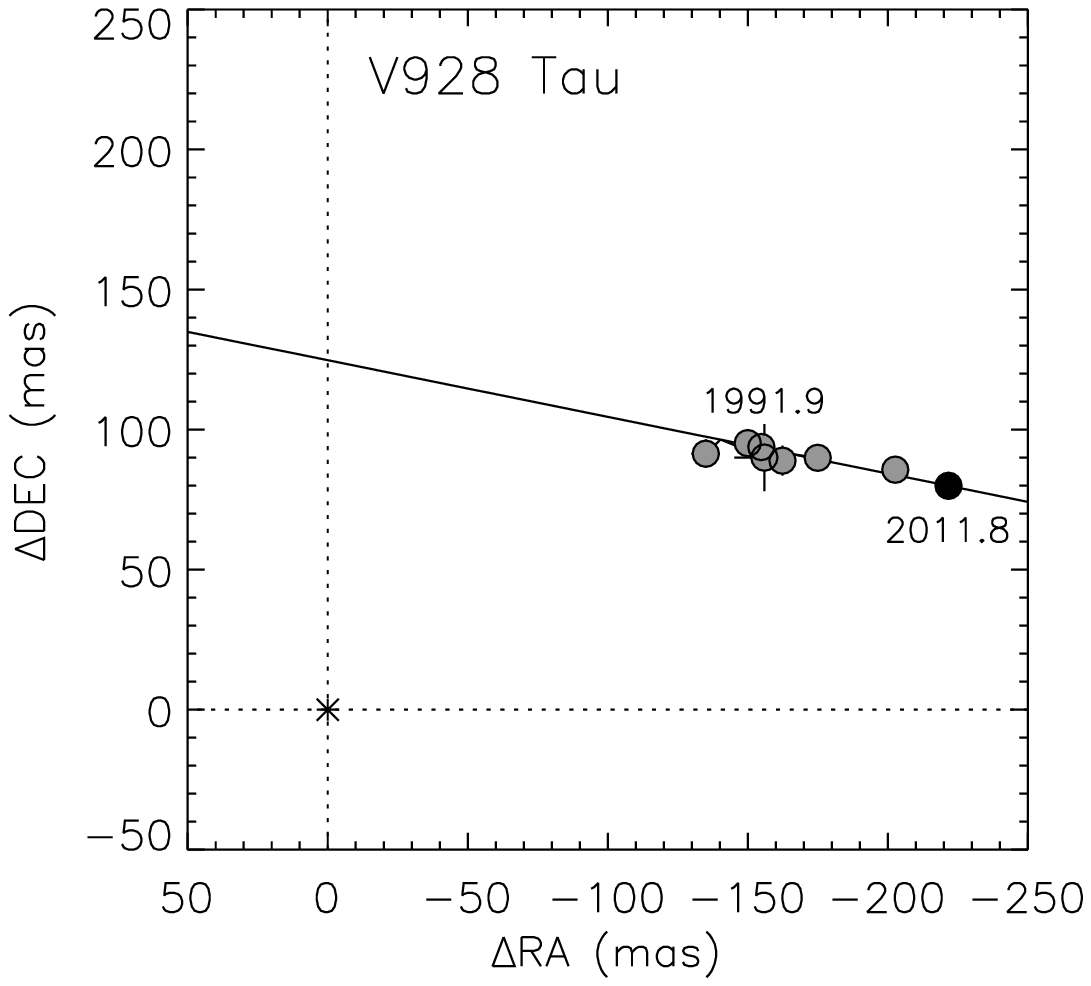}}
	\caption{Orbital motion measured for V928 Tau.  The black circles are NIRC2 measurements presented in Table~\ref{tab.sepPA}.  The grey circles are published values from the literature \citep{leinert93, ghez95, simon96, white01, kraus12}.  The solid line shows a fit for linear motion.  If bound, then a statistical analysis of orbital solutions that fit the data indicates that $P > 58$ yr.  The position angles from \citet{ghez95}, \citet{simon96}, and \citet{white01} were flipped by 180$^\circ$ for consistency.  This is reasonable for the nearly equal brightness pair.}
\label{fig.v928tau_orb}
\end{figure}

{\it V928 Tau.} --  
V928 Tau is classified as a weak-lined T Tauri star with a spectral type of M0.5 \citep{martin94, sestito08}.  It is a binary system with two components of nearly equal brightness.  The separation of the binary ranges from 180 to 240 mas \citep[][; this work]{leinert93, ghez93, ghez95, simon96, white01, kraus12}.  Figure~\ref{fig.v928tau_orb} shows the orbital motion for V928 Tau based on our AO observations and previously published measurements.  Currently, the orbital motion is indistinguishable from linear motion.  If the pair is bound, then a statistical analysis of orbits that fit the data indicates that the orbital period must be greater than 58 yr.

\section{Discussion}
\label{sect.discussion}

\subsection{Are the linear motion binaries physical pairs?}
\label{sect.linear}

Of the 16 close binaries that were sampled (with separations ranging from 30 to 350 mas), five show relative motion that is indistinguishable from linear motion (GH Tau, Haro 6-37, HBC 360, IS Tau, V928 Tau).  The question then becomes, have we not observed these systems long enough to see curvature in their orbits or are they chance alignments with other cluster members or background stars?  Each of these possibilities is discussed further below.

The typical projected separation for stars in nearby star forming regions is $\sim$~20,000 AU \citep{reipurth93} which corresponds to $\sim 130''$ at 140 pc.  Assuming a random distribution of stars with this average separation, the probability of a chance alignment between two cluster members with a projected separation of 0.35$''$ ($\sim 50$ AU) would be $\lesssim 0.2$\%.  This suggests that the close binaries in our sample are highly likely to be physical pairs.

We can rule out the likelihood of chance alignments with field or background stars based on the relative proper motion of the components in the binaries.  The average proper motion for stars in the Taurus region is $\mu_\alpha \cos{\delta} = 7.16 \pm 8.55$ mas yr$^{-1}$ and $\mu_\delta = -20.91 \pm 10.31$ mas yr$^{-1}$ \citep{bertout06}.  Table~\ref{tab.pm} lists the proper motion measured by \citet{ducourant05} for FW Tau and each of the linear motion systems in our sample.  The last two columns of the Table give the relative motion of the secondary relative to the primary from the linear fit in RA and Dec.  Except for FW Tau, the motion of each component relative to its primary is small ($<$ 6 mas yr$^{-1}$) in comparison to the proper motion measured for the system.  All of the proper motions are below the cut-off value of $\Delta \mu < 12$ mas yr$^{-1}$ that \citet{kraus09} use to define comoving pairs within nearby star forming regions.  Both components in GH Tau show signs of accretion and are both identified as classical T Tauri stars \citep{white01,hartigan03}, providing strong evidence that they are members of the Taurus star forming region.  The components in FW Tau have a slightly larger relative proper motion compared with the other systems; however, as shown in Figure~\ref{fig.fwtau_orb}, the position measurements for this equal brightness binary can also be combined to show curvature rather than linear motion.

\begin{figure}
	\scalebox{1.0}{\includegraphics{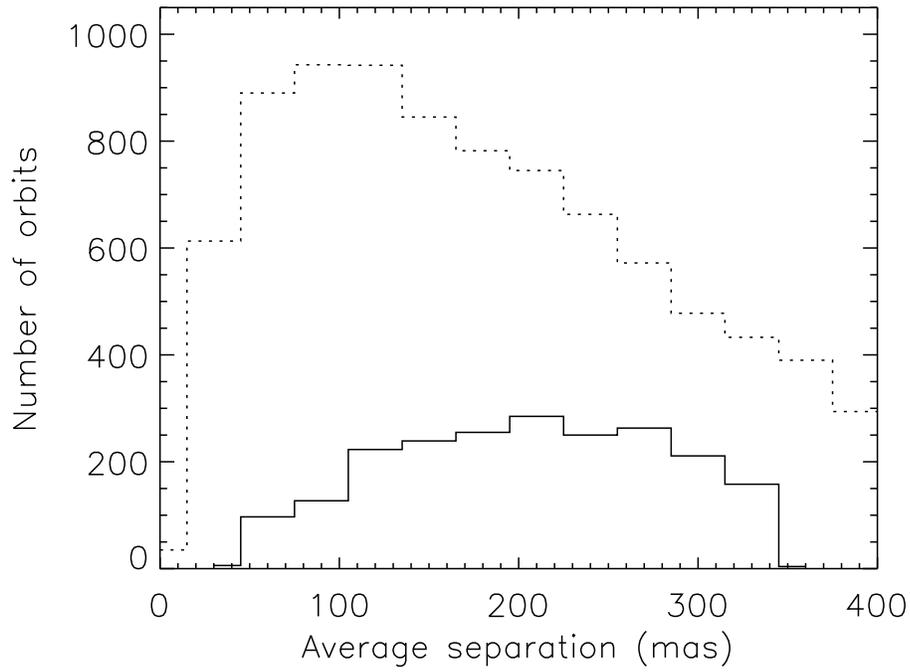}}
	\caption{Random sample of 10,000 binary orbits (see Section~\ref{sect.linear} for selection criterion).  The dotted line shows the distribution of average separations for these binaries measured over a 20 year time frame.  The solid line shows the distribution of those orbits where the measured separation is between 30$-$350 mas and the relative motion between the components is consistent with linear motion.}
\label{fig.linear}
\end{figure}

As a final check, we ran simulations to see how likely it is that a binary orbit could be resolved with separations as small as 30$-$350 mas yet show linear motion over the time-frame of the observations.  We randomly created a sample of 10,000 binary orbits with periods ranging from 5 to 500 yr, time of periastron passage within one orbital period of the year 2000, eccentricities ranging from 0 to 0.99, inclinations and $\Omega$ ranging from 0 to 180$^\circ$, and $\omega$ ranging from 0 to 360$^\circ$.  We set constraints on the semi-major axis by requiring that at a distance of 140 pc, the total mass of the binary had to be between 0.08~$M_\odot$ (set by hydrogen burning limit) and 2~$M_\odot$ (set by two 1~$M_\odot$ stars, consistent with the late spectral type stars in our sample) for a given period.   All parameters were drawn from uniform distributions.  For each orbit, we calculated the position of the binary at two year intervals from 1990 to 2010, roughly consistent with the time-frame over which most of the binaries were observed.  We added random Gaussian noise to the positions, with an uncertainty of 1~mas and calculated $\chi^2_{\rm orb}$ between the simulated measurements and the noise-free orbital positions.  We then computed linear fits to the motion in RA and Dec and computed $\chi^2_{\rm lin}$ between the simulated measurements and the linear fit.  We performed an F-test to determine whether the additional parameters in the orbital solution provided an improvement over the linear fit using a significance level of 0.05.  In our random sample of 10,000 orbits, we found that 21\% $\pm$ 4\% of the orbits had measured separations between 30$-$350 mas and showed motion consistent with linear motion over a 20 yr time-frame.  This is not significantly different from our observed sample where 31\% $\pm$ 12\% of the binaries show linear motion.  Figure~\ref{fig.linear} shows how the simulated distribution of separations of the linear motion binaries compares with the full random sample.  The probability of finding a binary with separations as small as 50$-$70 mas that shows linear motion over 20 yr drops off significantly.  We measure separations this small for HBC 360, however, for that system we only have measurements that span 7 yrs, so the probability of finding an orbit that shows linear motion increases dramatically for that case.  Overall, these simulations suggest that there is a high likelihood that we simply have not waited long enough for the orbital motion for the five linear systems to show curvature.  Continued observations will confirm whether these systems are physical pairs.

\subsection{Comparison of $M_{\rm tot}$ with evolutionary tracks}

Ultimately, our goal of determining precise masses of PMS stars is to compare with predictions from different sets of evolutionary models in order to determine which give the most accurate representation of PMS evolution.  The total masses for DF Tau and ZZ Tau in Table~\ref{tab.orbpar} are still preliminary and since we do not know their individual distances, we used the average distance of 140 pc to the Taurus star forming region in computing their masses.  Nonetheless, we plot the components on the H-R diagram to confirm that our dynamical masses make sense in an evolutionary context.

We assumed spectral types of M1V for DF Tau A and B and spectral types of M2.5V and M3.5V for ZZ Tau A and B, respectively.  These are based on spatially resolved, high resolution IR spectra of the components (Prato et al., in preparation).  We estimated effective temperatures based on these spectral types using the temperature scale adopted by \citet{luhman03} and assumed uncertainties of $\pm$ 1 spectral subtype.

In Table~\ref{tab.mag} we present average magnitudes for the components in DF Tau and ZZ Tau.  The $V$-band magnitudes represent an average of the calibrated magnitudes measured with the FGS in \citet{schaefer03,schaefer06}.  To determine the infrared magnitudes, we computed the average AO flux ratio between the components in each filter from Table~\ref{tab.sepPA}.  We combined these flux ratios with the total $JHK$ magnitudes from 2MASS \citep{cutri03} and the total $L$-band magnitudes from \citet{kenyon95} to compute magnitudes for the individual components.  \citet{white01} also measured $UBVRI$ magnitudes of the components in DF Tau using {\it HST}.  

\begin{figure}
	\scalebox{0.62}{\includegraphics{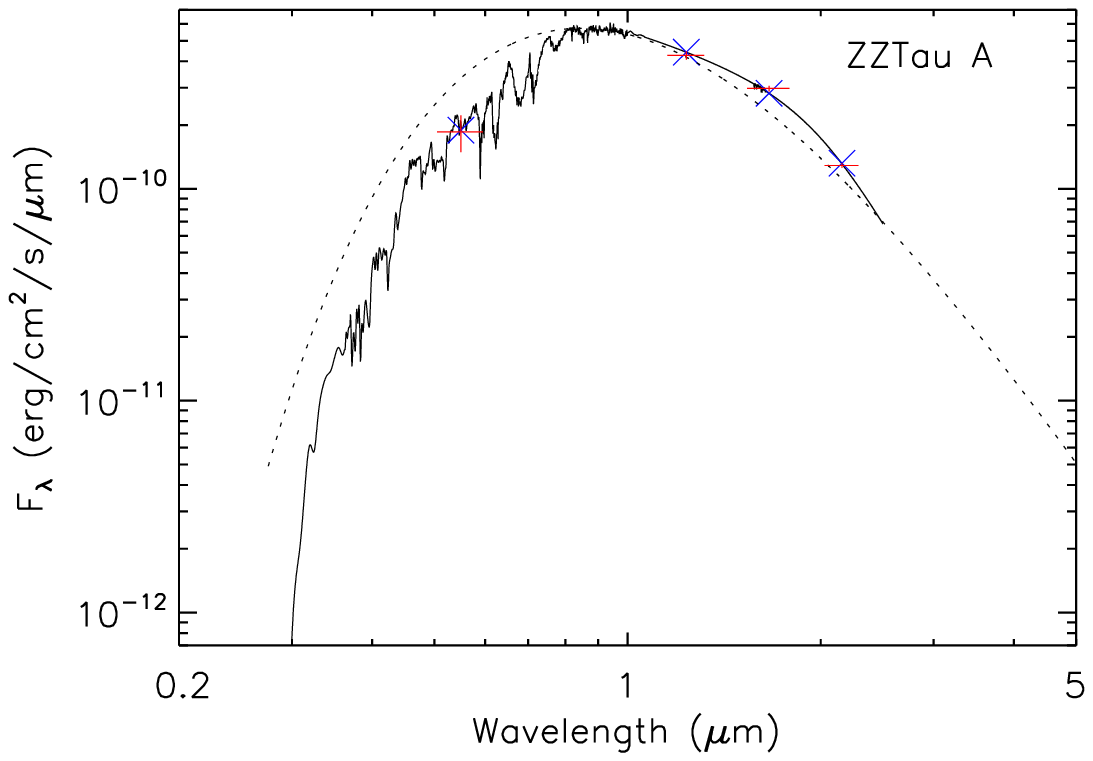}}
	\scalebox{0.62}{\includegraphics{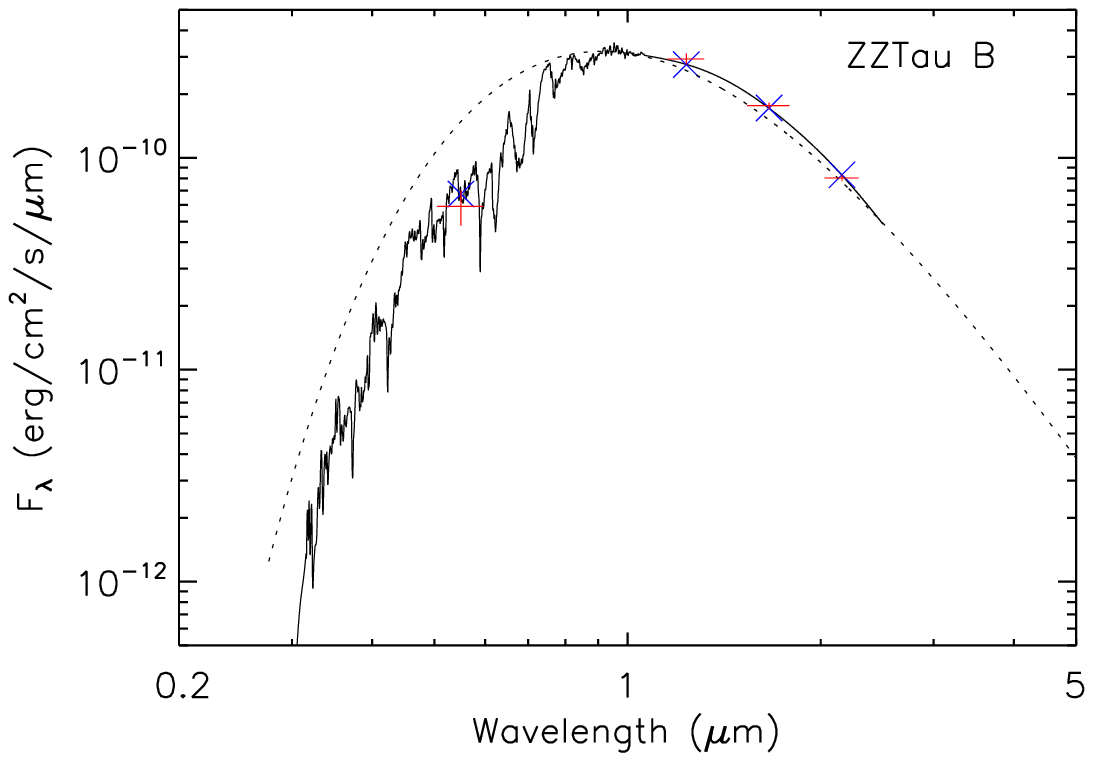}}
        \caption{Spectral energy distributions for ZZ Tau A and B.  We fit the photometric measurements using the Pickles spectral templates at M2.5V and M3.5V for A and B, respectively.   The M3.5 template was interpolated between the M3 and M4 templates.  We use the Rayleigh Jeans tail of a blackbody curve to approximate the flux at wavelengths longer than 2.5$\mu$m (dotted line).  The average model flux across the given photometric band is shown by the blue crosses.  The red plus symbols show the measured photometric magnitudes; the vertical length gives the size of the measurement error in the magnitudes while the horizontal width represents the width of the photometric filter.}
\label{fig.sed}
\end{figure}

For DF Tau, we used the luminosities of the components computed by \citet{hillenbrand04}.  For ZZ Tau A and B, we computed luminosities by constructing spectral energy distributions (SED) using the magnitudes listed in Table~\ref{tab.mag}.    We converted the observed magnitudes $m_\lambda$ to measured fluxes through 
\begin{equation}
F_{\lambda, \rm meas} = F_{\lambda_0} \times 10.0^{-0.4[m_\lambda - A_\lambda]}
\end{equation}
where $A_\lambda$ is the wavelength-dependent interstellar extinction assuming an R=3.1 reddening law \citep{cardelli89,odonnell94} and $F_{\lambda_0}$ is the zero-point flux density for a filter at wavelength $\lambda$.  We fit the flux calibrated spectral templates from \citet{pickles98} to the measured photometry by averaging the flux of the spectral templates across the width of each of the observed filter bands.  The emergent flux from the star given by the templates scales as the area subtended on the sky (or as the square of the stellar angular diameter).  Therefore, fitting the SED with stellar templates is dependent on three parameters, the spectral type of the template, the angular diameter of the star $\theta$, and the interstellar extinction $A_V$.  We computed the bolometric flux ($F_{\rm bol}$) by summing the flux underneath the best fitting SED.  The Pickles spectral templates only extend out to 2.5$\mu$m, so we used the Rayleigh Jeans tail of a blackbody curve to approximate the flux at longer wavelengths.  The luminosity is computed from $L = 4 \pi d^2 F_{\rm bol}$, assuming a distance of $d$ = 140 pc.  The SED fits for ZZ Tau A and B are shown in Figure~\ref{fig.sed}.  We fixed the spectral type of the templates  and computed uncertainties in $A_V$ and $L$ by varying the template by $\pm 1$ spectral subtype and recomputing the SED fit.

\begin{figure}
\begin{center}
	\scalebox{0.62}{\includegraphics{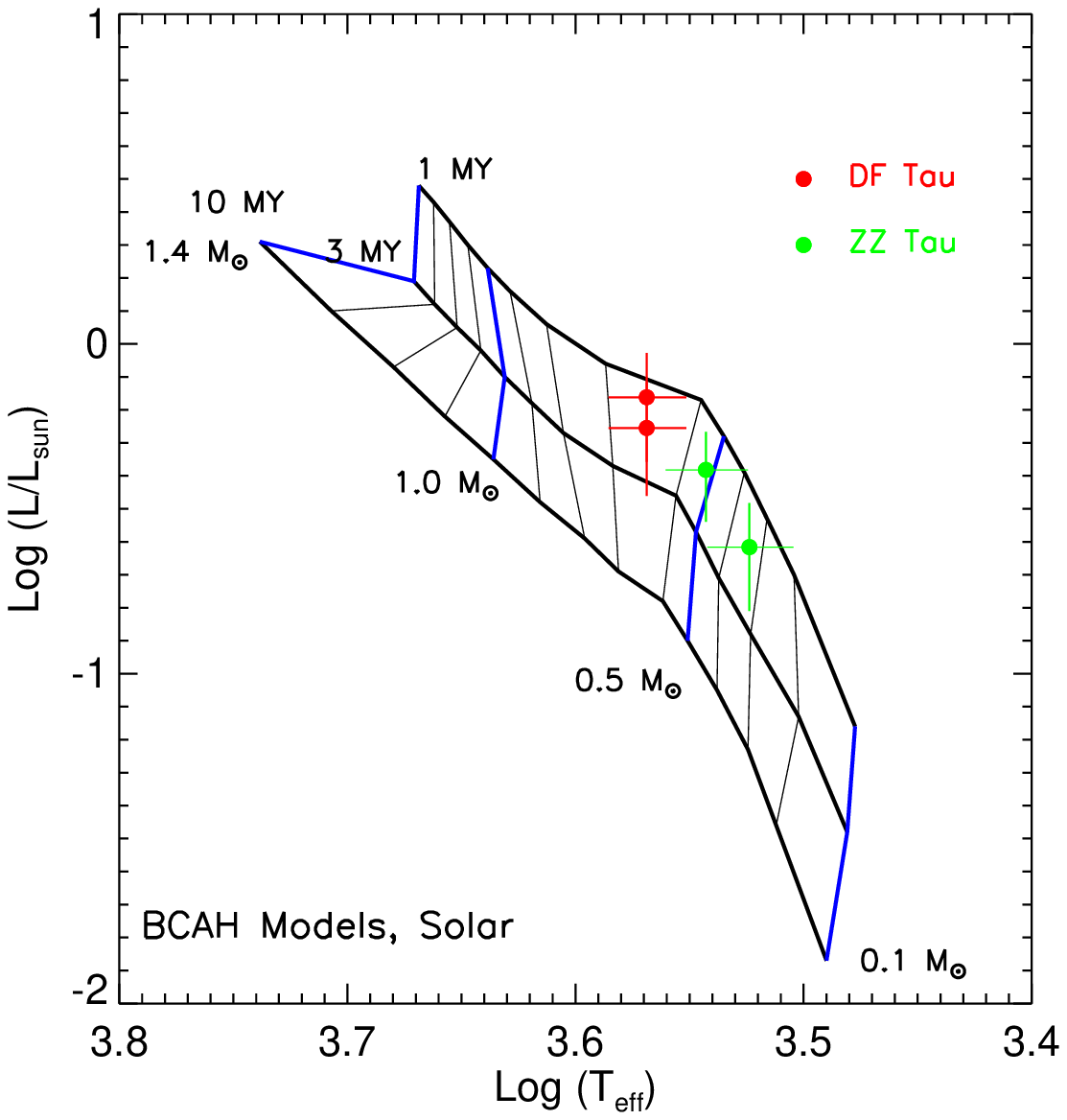}}
	\scalebox{0.62}{\includegraphics{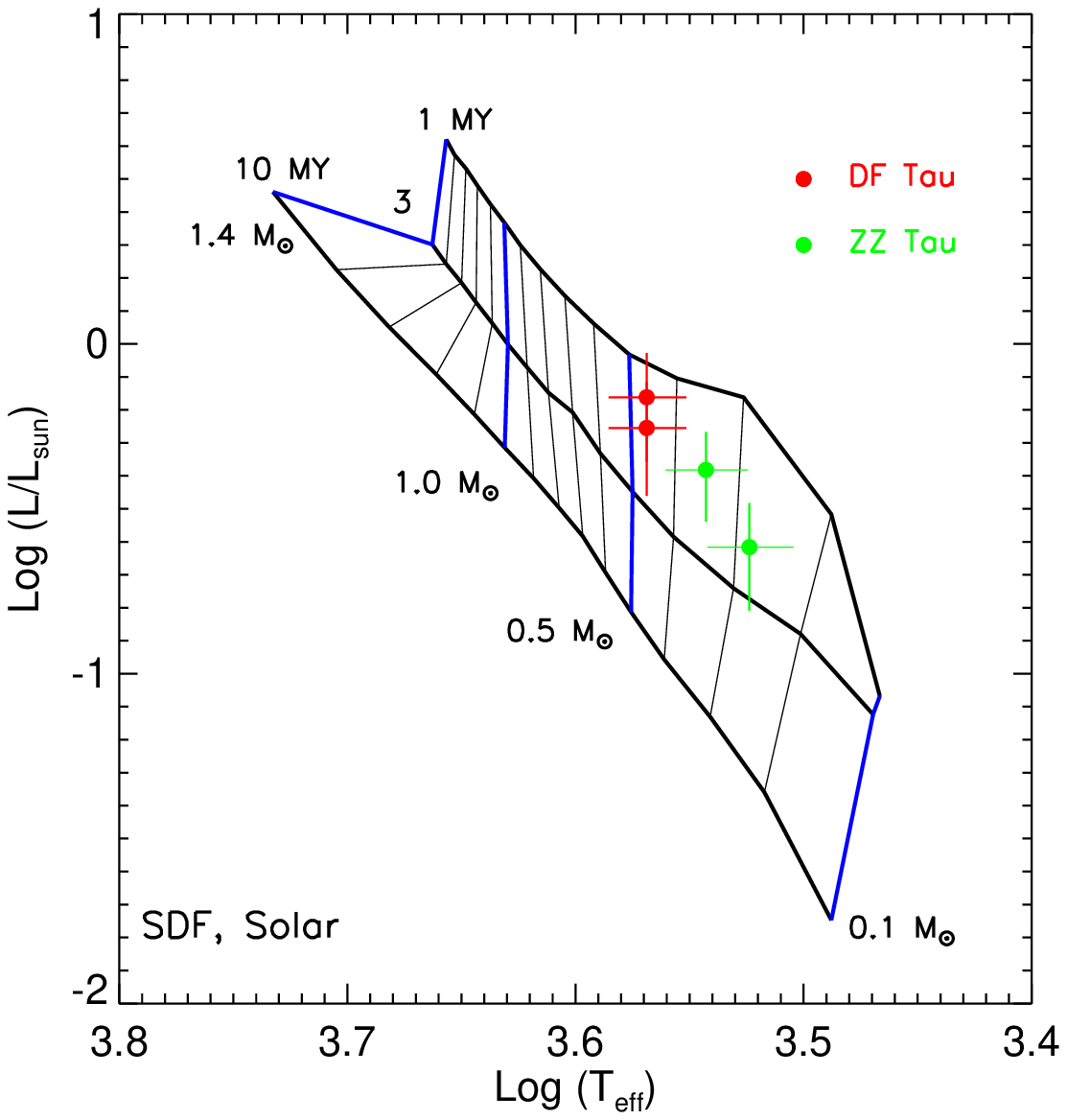}}
	\scalebox{0.62}{\includegraphics{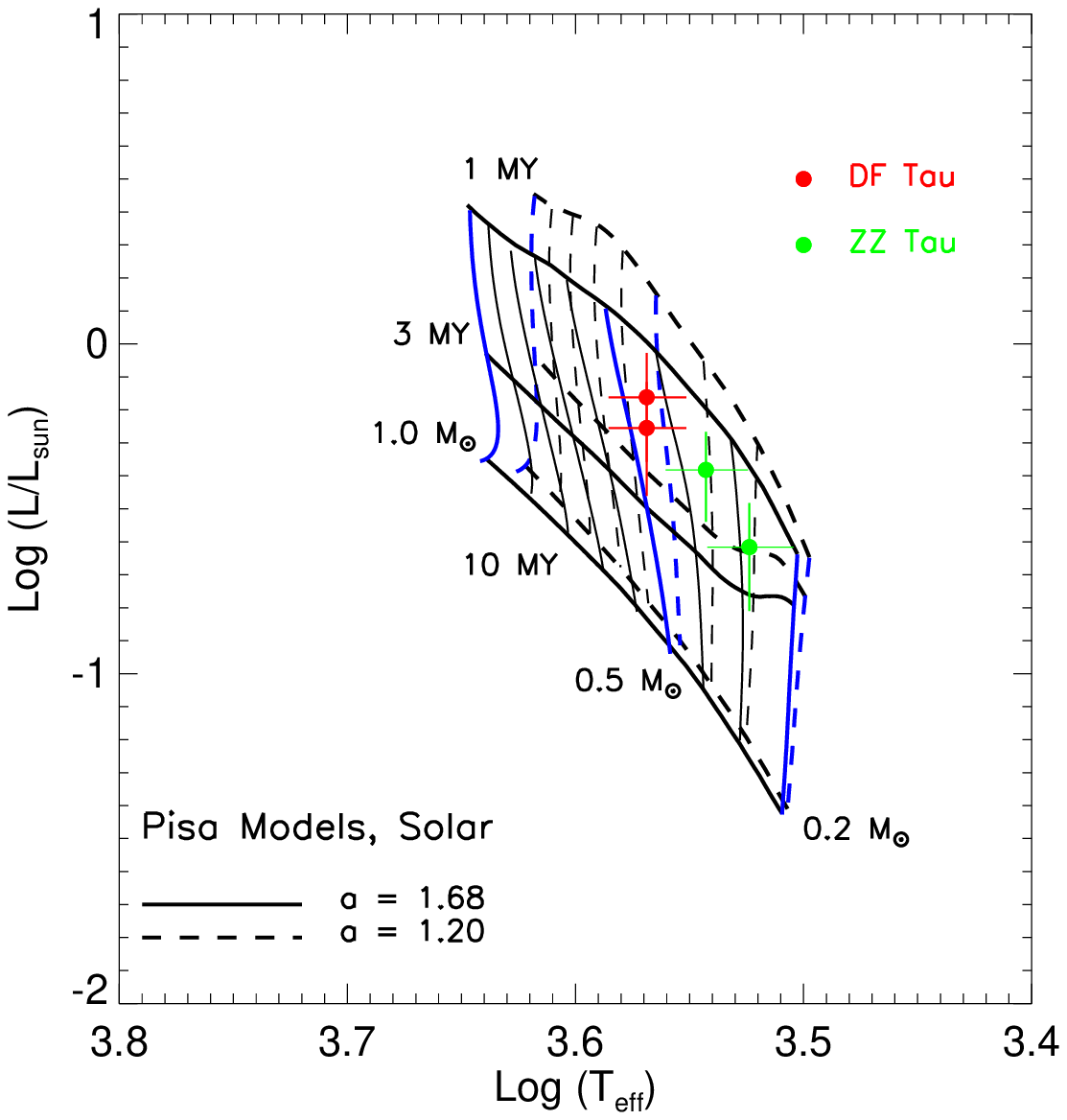}}
	\scalebox{0.62}{\includegraphics{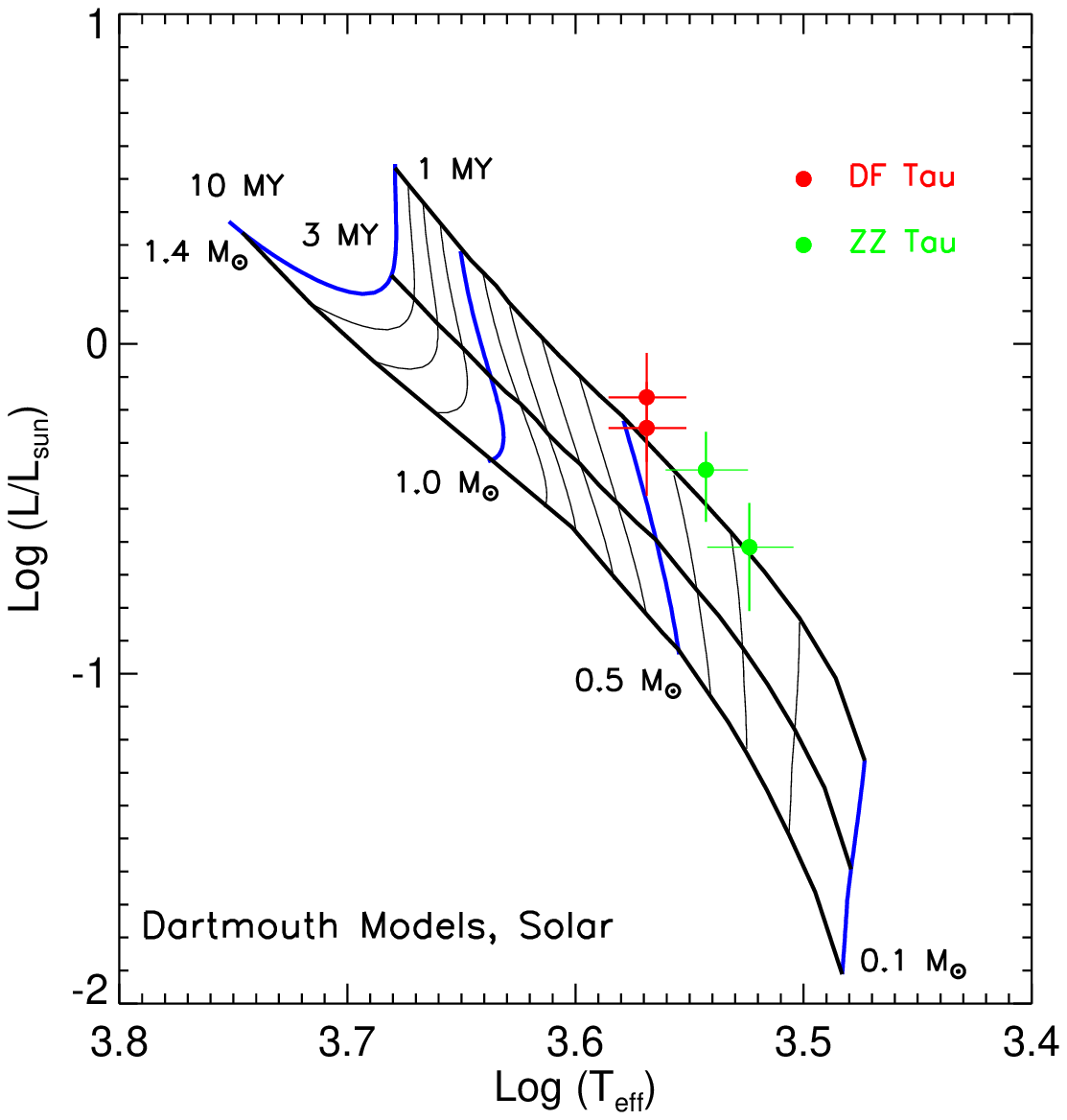}}
\end{center}
	\caption{H-R diagrams showing the location of DF Tau A-B and ZZ Tau A-B.  {\it Upper Left:} Evolutionary tracks of BCAH (Baraffe et al.\ 1998) at solar metallicity for masses of 0.1 to 1.4 M$_\odot$.  For masses below 0.7 M$_\odot$, we used the tracks calculated with a mixing length parameter of 1.0; for larger masses we used the tracks with a mixing length of 1.9.  {\it Upper Right:} Evolutionary tracks of SDF (Siess et al.\ 2000) at $Z=0.02$ for masses of 0.1 to 1.6 M$_\odot$.  {\it Lower Left:} Pisa stellar evolutionary models (Tognelli et al.\ 2011) for mixing length parameter of $\alpha = 1.20$ (solid lines) and $\alpha = 1.68$ (dashed lines) at a metallicity of $Z=0.02$.  The evolutionary tracks are plotted from 0.2 to 1.0 M$_\odot$ at 0.1 M$_\odot$ intervals. {\it Lower Right:} Dartmouth evolutionary models (Dotter et al. 2008) at solar metalicity for masses of 0.1 to 1.4 M$_\odot$.  In all of the panels, the isochrones are plotted at 1, 3, and 10 Myr.  The location of DF Tau A-B and ZZ Tau A-B are marked by the red and green circles, respectively.  The error bars on the luminosities include $\pm$~20~pc uncertainty on the distance and those for the effective temperatures assume  uncertainties of $\pm$~1 spectral subtype.}
\label{fig.tracks}
\end{figure}

Table~\ref{tab.properties} summarizes the physical properties we adopted for the components in DF Tau and ZZ Tau.  We do not include the distance uncertainty in the luminosity estimates for ZZ Tau in the Table.  In Figure~\ref{fig.tracks}, we plot the components of DF Tau and ZZ Tau in the H-R diagram relative to the evolutionary tracks computed by \citet[][, BCAH]{baraffe98}, \citet[][, SDF]{siess00}, \citet[][, Pisa]{tognelli11}, and \citet[][, Dartmouth]{dotter08}.  For all sets of tracks, the sum of the evolutionary masses of the A and B components in each system agrees at the 1\,$\sigma$ level with the total dynamical mass determined from our preliminary orbit fit.  The BCAH, SDF, and Pisa tracks suggest a young age of $2 \pm 1$~Myr for DF Tau and ZZ Tau, while the Dartmouth tracks tend to prefer a slightly younger age of $\lesssim$ 1 Myr.  Although \citet{hillenbrand04} use $I_C$-band magnitudes in determining the luminosities in order to minimize the impact of continuum excess from a circumstellar disk, the placement of DF Tau on the H-R diagram could be affected by its photometric variability \citep[e.g.,][]{wolk13}.


A more rigorous comparison with the evolutionary tracks requires improving the precision in our dynamical mass measurements and refining our estimates of the component effective temperatures.  The improvement in $M_{\rm tot}$ can be accomplished, in part, with more complete coverage of the binary orbits \citep[e.g.,][]{schaefer12}.  However, it also requires a more precise distance measurement to each system.  Although the orbital periods are long, we are attempting to measure the radial velocity variations of the binaries in our sample using NIRSPAO at the Keck Observatory.  Combined with the visual orbit, this would provide individual masses and the orbital parallax.  As we show in Figure~\ref{fig.taurus}, it might be possible to infer the distance to some binaries by their association within particular clouds that have a member with a precise distance measurement.  For instance, DF Tau and ZZ Tau lie within the same region as V807 Tau, for which we are measuring a parallax using the VLBA \citep{schaefer12}.  In the future, GAIA will provide high precision parallaxes for more stars in the Taurus region.  Improvements in the effective temperatures could be accomplished through a direct comparison of high-resolution spectra with atmospheric models \citep[e.g.,][]{rice10} or by assigning precise temperatures to the observed spectral templates rather than using a spectral type conversion table \citep{torres13}.

\section{Summary}
\label{sect.summary}

We present updated orbital measurements of 16 binary and multiple systems using adaptive optics imaging at the Keck Observatory.  Together with previously published measurements available in the literature, we analyzed the current orbital motion for each system.  Our results are as follows:

1. We determined preliminary orbital parameters for four binaries: DF Tau, T Tau S, ZZ Tau, and the Pleiades binary HBC 351.  The preliminary orbits represent a reasonable description of their motion in the near future, although the parameters are likely to be revised with additional measurements in the longer-term.  The total masses for these systems are consistent with predictions from evolutionary tracks.  Additional orbital measurements to improve the precision of the masses will allow a more rigorous comparison between different sets of evolutionary tracks.

2. Seven binaries show curvature in their relative motion, but we can only place lower limits on their periods based on the limited orbital coverage (FO Tau, FW Tau, HBC 361, HV Tau, RX J0435.9+2352, UX Tau B, V410 Tau). Full solutions will be possible with more complete orbital coverage.

3. Five binaries exhibit motion that is indistinguishable from linear motion (Haro 6-37, HBC 360, GH Tau, IS Tau, V928 Tau).  The small relative motion between the components ($<$ 6 mas\,yr$^{-1}$) suggests that these systems are bound.  Additionally, simulations suggest that $\sim$ 20\% of orbits drawn randomly from a sample ($P = 5-500$ yr, $M_{\rm tot} = 0.08-2~M_\odot$) will have separations between $30-350$ mas and show linear motion over a time-frame of 20~yr.  Additional observations will confirm whether these systems are bound.

\acknowledgements

\begin{figure}
	\scalebox{1.0}{\includegraphics{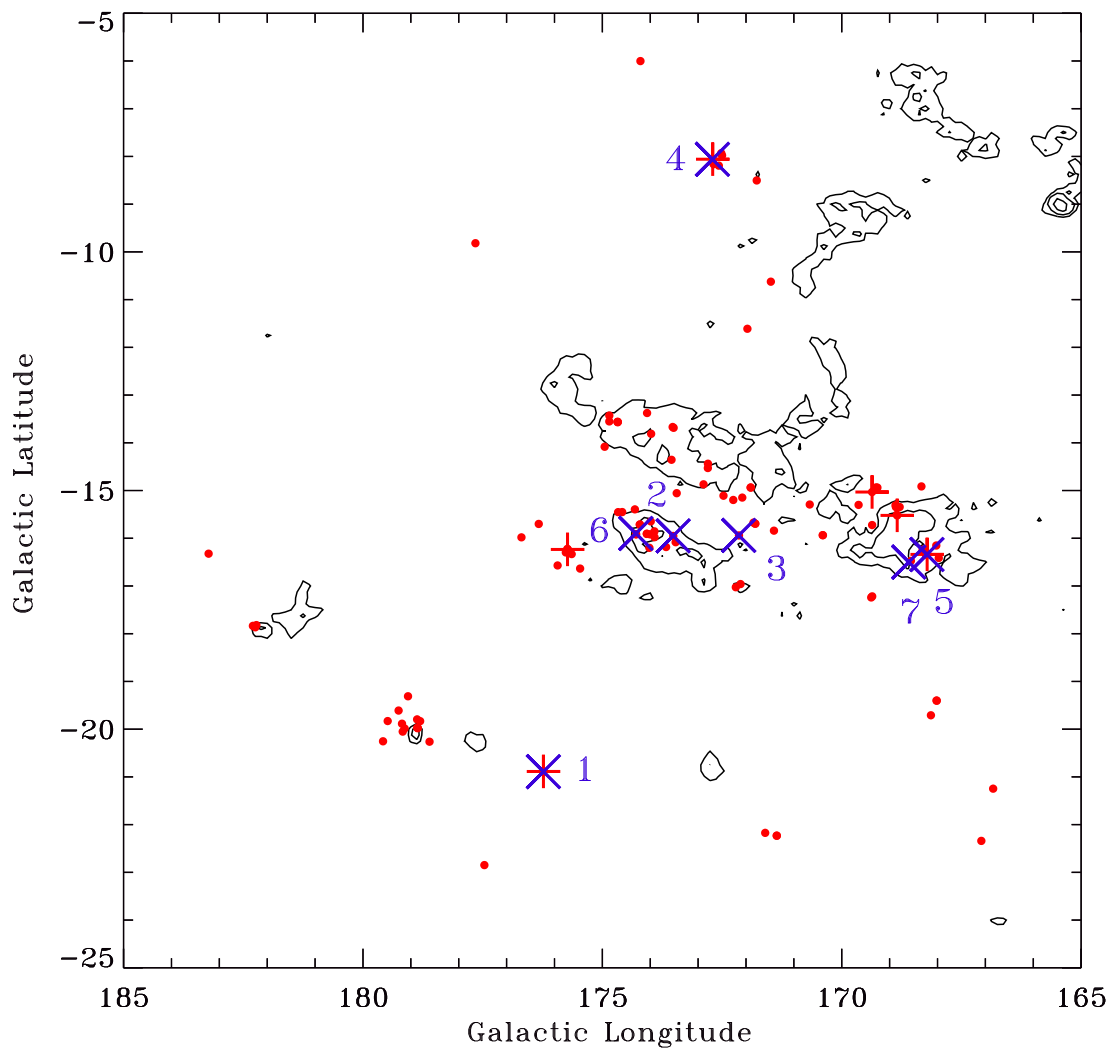}}
	\caption{Spatial distribution of stars in the Taurus star forming region.  The blue X's are binaries with enough orbital coverage to estimate their total mass: 1) T Tau, 2) ZZ Tau, 3) DF Tau, 4) NTTS 045251+3016, 5) V773 Tau, 6) V807 Tau, and 7) LkCa3.  The red crosses are stars with precises distances \citep[e.g.,][]{torres09}.  The small red dots show the location of other stars in Taurus that have measured proper motions \citep{ducourant05,kenyon08}.  The contours outline the integrated CO intensities \citep{dame01}.}
\label{fig.taurus}
\end{figure}

We thank the staff at Keck Observatory for their excellent support during our NIRC2 observing runs.  We thank J.\ Stern for making Figure~\ref{fig.taurus}.  Data presented herein were obtained at the W. M. Keck Observatory from telescope time allocated to the National Aeronautics and Space Administration through the agency's scientific partnership with the California Institute of Technology and the University of California.  Keck telescope time was also granted by NOAO, through the Telescope System Instrumentation Program (TSIP). TSIP is funded by NSF. The Observatory was made possible by the generous financial support of the W. M. Keck Foundation.  We recognize the Hawaiian community for the opportunity to conduct these observations from the summit of Mauna Kea.  GHS acknowledges support from NASA Keck PI Data Awards administered by the NASA Exoplanet Science Institute (JPL contracts 1441975 and 1467049).  LP acknowledges support from NSF grant AST-1009136; MS acknowledges support from NSF grant AST-09-08406.  This research has made use of the SIMBAD database and the VizieR catalog access tool, CDS, Strasbourg, France.

\clearpage


\clearpage

\begin{deluxetable}{llllclll}
\tabletypesize{\scriptsize}
\tablecaption{Log of NIRC2+AO Observations\tablenotemark{a}}
\tablewidth{0pt}
\tablehead{
\colhead{UT Date} & \colhead{UT} &\colhead{Target}&\colhead{Filter} & \colhead{AO Rate} & \colhead{$T_{\rm int}$ (s)\tablenotemark{b}} &\colhead{No. Images\tablenotemark{c}} &\colhead{PSF Used}}
\startdata
2006 Dec 06  & 10:43 & HBC 351    &  Kcont                     & 248 & 1.0          &  4         & HBC 351 C       \\
             & 12:38 & HBC 360    &  Kcont                     &  55 & 1.0          &  4         & HBC 365         \\
             & 12:53 & HBC 361    &  Kcont                     &  55 & 1.0          &  4         & HBC 365         \\
             & 13:10 & HBC 365    &  Kcont                     & 125 & 1.0          &  4         & PSF             \\
2006 Dec 18  & 05:15 & T Tau      &  Kcont                     & 660 & 0.18         & 10         & T Tau N         \\
             & 06:12 & DF Tau     &  Jcont, Hcont, Kcont       & 338 & 0.18 -- 0.5  & 10, 10, 10 & DN Tau          \\
             & 06:31 & DN Tau     &  Jcont, Hcont, Kcont       & 272 & 0.5 -- 1.0   & 5, 5, 5    & PSF             \\
             & 06:47 & DN Tau     &  J, H, K$'$                & 272 & 0.18 -- 0.4  & 5, 5, 5    & PSF             \\
             & 07:07 & ZZ Tau     &  J, H, K$'$                & 114 & 0.4 -- 0.6   & 10, 10, 10 & DN Tau          \\
2008 Jan 17  & 05:25 & DF Tau     &  Kcont, L$'$               & 438 & 0.05 -- 0.18 & 10, 10     & DN Tau          \\
             & 05:41 & DN Tau     &  Kcont, L$'$               & 438 & 0.18 -- 0.5  & 10, 10     & PSF             \\
             & 05:56 & ZZ Tau     &  Kcont, L$'$               & 149 & 0.18 -- 0.5  & 10, 10     & DN Tau          \\  
             & 07:16 & HIP 23418\tablenotemark{d} & Jcont, Hcont, Kcont, L$'$ & 438 & 0.05 -- 0.18 & 10, 10, 10, 10 & PSF, 1$\arcsec$ bin \\
             & 07:42 & HV Tau     &  Jcont, Hcont, Kcont, L$'$ & 149 & 0.18 -- 1.0  & 10, 10, 10, 10 & HIP 23418 A  \\
             & 08:36 & T Tau      &  Kcont                     &1054 & 0.18         & 10         & T Tau N          \\
             & 09:19 & HBC 360    &  Kcont                     & 149 & 5.0          & 10         & DN Tau           \\
             & 09:33 & HBC 361    &  Kcont                     & 149 & 5.0          & 10         & DN Tau           \\
             & 09:46 & DN Tau     &  Kcont                     & 438 & 0.5          & 10         & PSF              \\
             & 09:54 & FF Tau     &  Kcont                     & 149 & 1.0          & 10         & unresolved       \\
2008 Dec 17  & 05:09 & T Tau      &  Kcont                     &1054 & 0.18         & 12         & T Tau N          \\
             & 05:55 & ZZ Tau     &  Hcont, Kcont              & 149 & 0.5, 1.0     & 12, 12     & DN Tau           \\
             & 06:12 & DN Tau     &  Hcont, Kcont              & 438 & 0.18         & 12, 12     & PSF              \\
             & 06:25 & DF Tau     &  Hcont, Kcont              & 438 & 0.4          & 6, 12      & DN Tau           \\
             & 07:27 & Haro 6-37  &  Kcont, L$'$               & 250 & 1.0, 0.18    & 12, 12     & Haro 6-37 A      \\
             & 07:55 & GH Tau     &  Kcont, L$'$               & 438 & 0.5, 0.18    & 12, 12     & GH Tau A         \\
             & 08:07 & IS Tau     &  Kcont, L$'$               & 149 & 0.18 -- 1.0  & 12, 12     & IS Tau A         \\
             & 08:41 & ZZ Tau     &  Kcont                     & 149 & 1.0          & 12         & DN Tau           \\
             & 08:54 & DN Tau     &  Kcont                     & 438 & 0.3          & 12         & PSF              \\
             & 09:29 & DN Tau     &  Hcont, Kcont              & 250 & 0.3 -- 1.0   & 12, 12     & PSF              \\
             & 10:03 & RXJ0435.9+2352 &  Hcont, Kcont          & 250 & 1.0          & 6, 12      & DN Tau           \\
2009 Oct 25  & 11:20 & T Tau      &  Kcont                     &1054 & 0.18         & 12         & T Tau N          \\
             & 11:30 & UX Tau     &  Kcont                     & 438 & 0.18 -- 0.5  & 8          & UX Tau A         \\
             & 11:52 & FW Tau     &  Kcont                     & 149 & 1.0 -- 2.0   & 15         & DN Tau           \\
             & 12:00 & DN Tau     &  Kcont                     & 438 & 0.18         & 6          & PSF              \\
             & 12:28 & HBC 352    &  H, K$'$                   & 438 & 0.18         & 6, 6       & PSF              \\
             & 12:40 & HBC 351    &  H, K$'$                   & 438 & 0.18         & 12, 12     & HBC 351 C        \\
             & 13:10 & HBC 360/361&  H, K$'$                   & 149 & 0.18         & 10, 10     & HBC 352          \\
             & 13:41 & HBC 352    &  H, K$'$                   & 149 & 0.18         & 3, 3       & PSF              \\
2011 Jan 24  & 05:10 & T Tau      &  Hcont, Kcont              &1054 & 0.18 -- 2.0  & 12, 9      & T Tau N          \\
             & 05:49 & DF Tau     &  Hcont, Kcont              & 438 & 0.5          & 6, 6       & DN Tau           \\
             & 06:08 & DN Tau     &  Hcont, Kcont              & 438 & 0.5          & 6, 6       & PSF              \\
             & 06:20 & ZZ Tau     &  Hcont, Kcont              & 438 & 0.5          & 12, 6      & DN Tau           \\
             & 06:34 & HBC 351    &  H, K$'$                   & 438 & 0.3          & 6, 6       & HBC 351 C        \\
             & 06:44 & HBC 352    &  H, K$'$                   & 438 & 0.18         & 6, 6       & PSF              \\
             & 07:06 & HBC 360/361&  H, K$'$                   & 149 & 0.5          & 10, 8      & HBC 352          \\
             & 07:21 & FW Tau     &  Hcont, Kcont              & 149 & 1.0 -- 2.0   & 12, 6      & DN Tau           \\
             & 07:50 & DN Tau     &  Hcont, Kcont              & 149 & 1.0          & 6, 6       & PSF              \\
             & 08:03 & HV Tau     &  Hcont, Kcont              & 149 & 0.3 -- 1.0   & 6, 6       & DN Tau           \\
2011 Oct 12  & 09:57 & UX Tau     &  Kcont                     & 250 & 0.5          & 12         & UX Tau A         \\
             & 10:11 & V410 Tau   &  Kcont                     & 438 & 0.5          & 12         & V410 Tau A       \\
             & 10:20 & FO Tau     &  Kcont                     & 149 & 1.0          & 12         & DN Tau           \\
             & 10:29 & DN Tau     &  Kcont                     & 149 & 0.5          & 12         & PSF              \\
             & 10:38 & V928 Tau   &  Kcont                     & 250 & 1.0          & 12         & V928 Tau A       \\
             & 10:47 & DN Tau     &  Kcont                     & 250 & 1.0          & 12         & PSF              \\
             & 10:56 & RXJ0435.9+2352&  Kcont                  & 250 & 1.0          & 12         & DN Tau           \\
             & 11:12 & T Tau      &  Hcont, Kcont              &1054 & 0.18         & 6, 18      & T Tau N          \\  
             & 11:53 & ZZ Tau     &  Hcont, Kcont              & 149 & 1.0          & 12 ,12     & DN Tau           \\
             & 12:09 & DN Tau     &  Hcont, Kcont              & 149 & 0.5          & 12, 12     & PSF              \\
             & 12:24 & DF Tau     &  Hcont, Kcont              & 149 & 0.18         & 12, 12     & DN Tau           \\
             & 12:37 & HBC 351    &  H, K$'$                   & 438 & 0.18         & 12, 12     & HBC 351 C        \\
             & 13:01 & HBC 360/361&  H, K$'$                   & 149 & 0.18 -- 0.5  & 14, 14     & HBC 352          \\
             & 13:18 & HBC 352    &  H, K$'$                   & 149 & 0.18         & 12, 6      & PSF              \\
             & 13:36 & FW Tau     &  Hcont, Kcont              & 149 & 2.0          & 12, 12     & DN Tau           \\
             & 13:51 & DN Tau     &  Hcont, Kcont              & 149 & 0.5          & 6, 6       & PSF              \\
             & 14:03 & HV Tau     &  Hcont, Kcont              & 149 & 1.0          & 12, 12     & DN Tau           \\
2013 Jan 27  & 05:03 & UX Tau     &  Kcont                     & 438 & 0.18 -- 0.5  & 12         & UX Tau A         \\
             & 05:15 & T Tau      &  Hcont, Kcont              &1054 & 0.18         & 6, 12      & T Tau N          \\
             & 05:27 & DF Tau     &  Hcont, Kcont              & 438 & 0.18         & 12, 12     & DN Tau           \\
             & 05:38 & DN Tau     &  Hcont, Kcont              & 438 & 0.5          & 6, 6       & PSF              \\
             & 05:53 & ZZ Tau     &  Hcont, Kcont              & 149 & 0.5 -- 2.0   & 12, 12     & DN Tau           \\
             & 06:07 & DN Tau     &  Hcont, Kcont              & 149 & 0.5 -- 1.0   & 6, 12      & PSF              \\
             & 06:22 & HBC 351    &  H, K$'$                   & 438 & 0.18         & 12, 12     & HBC 351 C        \\
             & 08:21 & FW Tau     &  Kcont                     & 149 & 2.0          & 12         & DN Tau           \\
             & 08:31 & DN Tau     &  Kcont                     & 149 & 1.0          & 6          & PSF              \\
             & 08:40 & FO Tau     &  Kcont                     & 149 & 2.0          & 12         & DN Tau           \\
             & 08:55 & DI Tau     &  Kcont                     & 438 & 1.0 -- 2.0   & 24         & comp not detected\\
             & 09:08 & DN Tau     &  Kcont                     & 438 & 1.0          & 12         & PSF              \\
             & 09:18 & V410 Tau   &  Kcont                     & 438 & 0.5          & 12         & V410 Tau A       \\
             & 09:27 & IS Tau     &  Kcont                     & 250 & 1.0          & 12         & DN Tau           \\
             & 09:39 & DN Tau     &  Kcont                     & 250 & 1.0          & 18         & PSF              \\
             & 09:53 & GH Tau     &  Kcont                     & 438 & 2.0          & 18         & GH Tau A         \\
             & 10:06 & Haro 6-37  &  Kcont                     & 438 & 2.0          & 12         & Haro 6-37 C      \\
2014 Jan 03  & 04:55 & HBC 351    &  H, K$'$                   & 438 & 0.18 -- 0.5  & 12, 12     & HBC 351 C        \\
             & 05:16 & HBC 360/361&  H, K$'$                   & 149 & 0.5          & 12, 10     & HBC 352          \\
             & 05:28 & HBC 352    &  H, K$'$                   & 149 & 0.18         & 6, 6       & PSF              \\
             & 05:37 & T Tau      &  Hcont, Kcont              &1054 & 0.18         & 6, 12      & T Tau N          \\
             & 06:19 & UX Tau     &  Hcont, Kcont              & 438 & 0.5  -- 0.6  & 12, 12     & UX Tau A         \\
2014 Jan 25  & 05:01 & DF Tau     &  Hcont, Kcont              & 438 & 0.5          & 12, 12     & DN Tau           \\
             & 05:12 & DN Tau     &  Hcont, Kcont              & 438 & 0.6          & 6, 6       & PSF              \\
             & 05:28 & RX J0435.9+2352&  Hcont, Kcont          & 438 & 1.0          & 12, 12     & DN Tau           \\
             & 05:44 & ZZ Tau     &  Hcont, Kcont              & 149 & 1.0          & 12, 12     & DN Tau           \\
             & 05:58 & DN Tau     &  Hcont, Kcont              & 149 & 1.0          & 6, 6       & PSF              \\
             & 06:09 & DN Tau     &  H, K$'$                   & 149 & 0.18         & 6, 6       & PSF              \\
             & 06:31 & FW Tau     &  H, K$'$                   & 149 & 0.8 -- 1.0   & 30, 30     & DN Tau           \\
\enddata
\tablenotetext{a}{See \citet{schaefer06} for a log of observations obtained prior to 2006.}
\tablenotetext{b}{Integration time per coadd.  Each image is composed of 10 coadded exposures, except for those in UT 2006 Dec 06 which are made up of 5 coadds.  If different exposure times were used for different sets of image, then the range of values is listed.}
\tablenotetext{c}{The number of images taken in each filter.}
\tablenotetext{d}{HIP 23418 is a member of $\beta$ Pic moving group and is a 1$\arcsec$ binary \citep{song03}.  We observed it as a PSF star.}
\label{tab.aolog}
\end{deluxetable}

\clearpage

\begin{deluxetable}{lllll} 
\tabletypesize{\scriptsize}
\tablewidth{0pt}
\tablecaption{NIRC2 Adaptive Optics Measurements of PMS Stars in Multiple Systems\tablenotemark{a}} 
\tablehead{
\colhead{Year} & \colhead{$\rho$(mas)} & \colhead{P.A.($\degr$)} & \colhead{Filter} & \colhead{Flux Ratio}}
\startdata 
\multicolumn{5}{c}{DF Tau A-B} \\
 \noalign{\vskip .8ex}%
 \hline
 \noalign{\vskip .8ex}%
2004.9815  &   109.93  $\pm$   0.49  &   246.06  $\pm$   0.26  & Kcont &  0.475 $\pm$ 0.014  \\
2005.9316  &   110.92  $\pm$   1.59  &   240.84  $\pm$   0.82  & K$'$  &  0.457 $\pm$ 0.028  \\
           &                         &                         & H     &  0.550 $\pm$ 0.023  \\
2006.9631  &   110.22  $\pm$   0.59  &   236.52  $\pm$   0.31  & Kcont &  0.624 $\pm$ 0.012  \\
           &                         &                         & Hcont &  0.827 $\pm$ 0.008  \\
           &                         &                         & Jcont &  0.914 $\pm$ 0.021  \\
2008.0445  &   109.97  $\pm$   0.35  &   231.57  $\pm$   0.18  & Kcont &  0.494 $\pm$ 0.012  \\
           &                         &                         & Lcont &  0.312 $\pm$ 0.005  \\
2008.9618  &   109.54  $\pm$   0.29  &   227.21  $\pm$   0.15  & Kcont &  0.535 $\pm$ 0.007  \\
           &                         &                         & Hcont &  0.818 $\pm$ 0.014  \\
2011.0645  &   106.88  $\pm$   0.35  &   217.24  $\pm$   0.19  & Kcont &  0.463 $\pm$ 0.006  \\
           &                         &                         & Hcont &  0.741 $\pm$ 0.010  \\
2011.7798  &   104.63  $\pm$   0.98  &   213.63  $\pm$   0.54  & Kcont &  0.450 $\pm$ 0.012  \\
           &                         &                         & Hcont &  0.733 $\pm$ 0.013  \\
2013.0740  &    99.97  $\pm$   1.35  &   206.84  $\pm$   0.78  & Kcont &  0.526 $\pm$ 0.018  \\
           &                         &                         & Hcont &  0.795 $\pm$ 0.015  \\
2014.0679  &    97.80  $\pm$   0.66  &   201.61  $\pm$   0.39  & Kcont &  0.503 $\pm$ 0.012  \\
           &                         &                         & Hcont &  0.797 $\pm$ 0.011  \\
\cutinhead{F0 Tau A-B}
2011.7796  &   137.22  $\pm$   0.17  &   241.10  $\pm$   0.07  & Kcont &  0.7394 $\pm$ 0.0082  \\
2013.0744  &   133.75  $\pm$   1.95  &   245.87  $\pm$   0.84  & Kcont &  0.8030 $\pm$ 0.0083  \\
\cutinhead{FW Tau A-B}
2009.8167  &    66.72  $\pm$   0.66  &   308.43  $\pm$   0.57  & Kcont &  0.994 $\pm$ 0.022  \\
2011.0646  &    78.73  $\pm$   0.83  &   311.52  $\pm$   0.60  & Kcont &  0.953 $\pm$ 0.013  \\
2011.7799  &    86.80  $\pm$   0.45  &   313.82  $\pm$   0.29  & Kcont &  0.954 $\pm$ 0.026  \\
           &                         &                         & Hcont &  0.997 $\pm$ 0.038  \\
2013.0744  &    92.93  $\pm$   1.63  &   314.17  $\pm$   1.01  & Kcont &  0.912 $\pm$ 0.042  \\
2014.0680  &   105.01  $\pm$   0.95  &   317.04  $\pm$   0.52  & K$'$  &  0.954 $\pm$ 0.018  \\
           &                         &                         & H     &  0.924 $\pm$ 0.011  \\
\cutinhead{FW Tau A-C}
2014.0680  &  2316.4   $\pm$   6.1   &   295.64  $\pm$   0.15  & K$'$  &  0.0051 $\pm$ 0.0004 \\
           &                         &                         & H     &  0.0027 $\pm$ 0.0003 \\
\cutinhead{GH Tau A-B}
2008.9620  &   299.26  $\pm$   0.19  &  288.497  $\pm$  0.037  & Kcont &  0.9302 $\pm$ 0.0081  \\
           &                         &                         & L$'$  &  0.9488 $\pm$ 0.0037  \\
2013.0746  &   294.69  $\pm$   1.10  &  286.229  $\pm$  0.213  & Kcont &  0.8971 $\pm$ 0.0242  \\
\cutinhead{Haro 6-37 A-B}
2008.9619  &   315.80  $\pm$   0.40  &   176.69  $\pm$   0.07  & Kcont &  0.1864 $\pm$ 0.0032  \\
           &                         &                         & L$'$  &  0.1979 $\pm$ 0.0043  \\
2013.0746  &   322.24  $\pm$  12.76  &   175.96  $\pm$   2.27  & Kcont &  0.1649 $\pm$ 0.0086  \\
\cutinhead{Haro 6-37 C-A}
2008.9619  &  2652.60  $\pm$   0.61  &   38.929  $\pm$  0.016  & Kcont &  0.492 $\pm$ 0.016  \\
           &                         &                         & L$'$  &  0.473 $\pm$ 0.019  \\
2013.0746  &  2644.36  $\pm$   8.32  &   38.929  $\pm$  0.180  & Kcont &  0.489 $\pm$ 0.011  \\
\cutinhead{HBC 351 A-B (member of Pleiades)}
2006.9308  &    54.21  $\pm$   4.66  &   299.90  $\pm$   4.93  & Kcont &  0.811 $\pm$ 0.051  \\
2009.8168  &    45.50  $\pm$   0.41  &   352.80  $\pm$   0.52  & K$'$  &  0.847 $\pm$ 0.010  \\
           &                         &                         & H     &  0.848 $\pm$ 0.016  \\
2011.0645  &    39.54  $\pm$   0.43  &    22.81  $\pm$   0.63  & K$'$  &  0.837 $\pm$ 0.018  \\
           &                         &                         & H     &  0.833 $\pm$ 0.011  \\
2011.7798  &    32.86  $\pm$   0.52  &    46.34  $\pm$   0.91  & K$'$  &  0.851 $\pm$ 0.012  \\
           &                         &                         & H     &  0.849 $\pm$ 0.007  \\
2013.0742  &    27.02  $\pm$   1.59  &   117.15  $\pm$   3.38  & K$'$  &  0.896 $\pm$ 0.109  \\
           &                         &                         & H     &  0.865 $\pm$ 0.065  \\
2014.0076  &    39.68  $\pm$   1.02  &   155.07  $\pm$   1.48  & K$'$  &  0.883 $\pm$ 0.005  \\
           &                         &                         & H     &  0.863 $\pm$ 0.018  \\
\cutinhead{HBC 351 C-A (member of Pleiades)}
2006.9308  &   562.06  $\pm$   2.52  &  134.195  $\pm$  0.257  & Kcont &  2.586 $\pm$ 0.223  \\ 
2009.8168  &   535.97  $\pm$   0.46  &  136.006  $\pm$  0.050  & K$'$  &  2.845 $\pm$ 0.027  \\ 
           &                         &                         & H     &  2.953 $\pm$ 0.019  \\ 
2011.0645  &   517.21  $\pm$   0.45  &  136.294  $\pm$  0.050  & K$'$  &  2.860 $\pm$ 0.031  \\ 
           &                         &                         & H     &  2.940 $\pm$ 0.015  \\ 
2011.7798  &   504.36  $\pm$   0.35  &  135.965  $\pm$  0.041  & K$'$  &  2.767 $\pm$ 0.025  \\ 
           &                         &                         & H     &  2.832 $\pm$ 0.031  \\ 
2013.0742  &   482.99  $\pm$   1.56  &  134.362  $\pm$  0.186  & K$'$  &  2.840 $\pm$ 0.178  \\ 
           &                         &                         & H     &  3.044 $\pm$ 0.133  \\ 
2014.0076  &   471.28  $\pm$   0.51  &  132.882  $\pm$  0.062  & K$'$  &  2.686 $\pm$ 0.021  \\
           &                         &                         & H     &  2.897 $\pm$ 0.068  \\
\cutinhead{HBC 351 C-B (member of Pleiades)}
2006.9308  &   509.70  $\pm$   3.04  &  135.700  $\pm$  0.341  & Kcont &  2.092 $\pm$ 0.134  \\ 
2009.8168  &   500.28  $\pm$   0.31  &  132.884  $\pm$  0.037  & K$'$  &  2.408 $\pm$ 0.017  \\ 
           &                         &                         & H     &  2.503 $\pm$ 0.042  \\ 
2011.0645  &   502.76  $\pm$   0.47  &  132.157  $\pm$  0.054  & K$'$  &  2.394 $\pm$ 0.035  \\ 
           &                         &                         & H     &  2.450 $\pm$ 0.035  \\ 
2011.7798  &   505.64  $\pm$   0.34  &  132.239  $\pm$  0.040  & K$'$  &  2.355 $\pm$ 0.027  \\ 
           &                         &                         & H     &  2.405 $\pm$ 0.034  \\ 
2013.0742  &   508.86  $\pm$   1.05  &  133.462  $\pm$  0.118  & K$'$  &  2.527 $\pm$ 0.146  \\ 
           &                         &                         & H     &  2.625 $\pm$ 0.115  \\ 
2014.0076  &   508.24  $\pm$   0.99  &  134.572  $\pm$  0.112  & K$'$  &  2.371 $\pm$ 0.013  \\
           &                         &                         & H     &  2.501 $\pm$ 0.033  \\
\cutinhead{HBC 360 A-B}
2006.9310  &    49.63  $\pm$   1.63  &   285.18  $\pm$   1.88  & Kcont &  0.981 $\pm$ 0.048  \\
2008.0449  &    54.37  $\pm$   0.63  &   290.56  $\pm$   0.66  & Kcont &  0.985 $\pm$ 0.003  \\
2009.8168  &    60.67  $\pm$   0.67  &   294.83  $\pm$   0.64  & K$'$  &  0.954 $\pm$ 0.030  \\
           &                         &                         & H     &  0.906 $\pm$ 0.062  \\
2011.0646  &    61.20  $\pm$   1.49  &   296.58  $\pm$   1.40  & K$'$  &  0.951 $\pm$ 0.026  \\
           &                         &                         & H     &  0.969 $\pm$ 0.081  \\
2011.7799  &    65.93  $\pm$   0.63  &   299.76  $\pm$   0.55  & K$'$  &  0.880 $\pm$ 0.021  \\
           &                         &                         & H     &  0.871 $\pm$ 0.034  \\  
2014.0077  &    71.09  $\pm$   0.48  &   302.70  $\pm$   0.39  & K$'$  &  0.990 $\pm$ 0.010  \\
           &                         &                         & H     &  0.938 $\pm$ 0.010  \\
\cutinhead{HBC 361 A-B}
2006.9310  &    63.98  $\pm$   1.11  &   252.57  $\pm$   0.99  & Kcont &  1.011 $\pm$ 0.054  \\
2008.0450  &    50.53  $\pm$   0.45  &   255.17  $\pm$   0.51  & Kcont &  0.985 $\pm$ 0.005  \\
2011.7799  &    29.61  $\pm$   2.83  &   228.45  $\pm$   5.48  & K$'$  &  0.992 $\pm$ 0.097  \\
           &                         &                         & H     &  0.801 $\pm$ 0.032  \\
2014.0077  &    69.78  $\pm$   0.57  &   245.90  $\pm$   0.47  & K$'$  &  0.998 $\pm$ 0.027  \\
           &                         &                         & H     &  0.968 $\pm$ 0.018  \\
\cutinhead{HBC 361 A relative to HBC 360 A}
2008.0450  &   7318.8  $\pm$    1.9  &   66.398  $\pm$  0.018  & Kcont &  1.110 $\pm$ 0.005  \\
2009.8168  &   7308.1  $\pm$    7.4  &   66.505  $\pm$  0.058  & K$'$  &  1.033 $\pm$ 0.093  \\
           &                         &                         & H     &  1.091 $\pm$ 0.175  \\
2011.0646  &   7313.1  $\pm$    11.8 &   66.260  $\pm$  0.093  & K$'$  &  0.633 $\pm$ 0.235  \\ 
           &                         &                         & H     &  0.758 $\pm$ 0.191  \\ 
2011.7799  &   7323.4  $\pm$    8.6  &   66.238  $\pm$  0.068  & K$'$  &  0.848 $\pm$ 0.072  \\
           &                         &                         & H     &  0.885 $\pm$ 0.075  \\
2014.0077  &   7337.3  $\pm$    6.5  &   66.292  $\pm$  0.052  & K$'$  &  0.800 $\pm$ 0.038  \\
           &                         &                         & H     &  0.754 $\pm$ 0.109  \\
\cutinhead{HIP 23418 = GJ 3322 (member of $\beta$ Pic moving group\tablenotemark{b})}
2008.0447  &  1295.80  $\pm$   0.62  &  153.117  $\pm$  0.029  & Kcont &  0.4917 $\pm$ 0.0082  \\
           &                         &                         & Hcont &  0.4972 $\pm$ 0.0066  \\
           &                         &                         & Jcont &  0.4958 $\pm$ 0.0110  \\
\cutinhead{HV Tau A-B}
2008.0448  &    40.89  $\pm$   0.95  &   323.25  $\pm$   1.33  & Kcont &  0.617 $\pm$ 0.007  \\
           &                         &                         & Hcont &  0.664 $\pm$ 0.016  \\
           &                         &                         & Jcont &  0.678 $\pm$ 0.068  \\
           &                         &                         & L$'$  &  0.616 $\pm$ 0.018  \\
2011.0647  &    25.83  $\pm$   2.04  &   331.68  $\pm$   4.52  & Kcont &  0.684 $\pm$ 0.150  \\
2011.7800  &    25.74  $\pm$   4.29  &   350.25  $\pm$   9.55  & Kcont &  0.308 $\pm$ 0.089  \\
           &                         &                         & Hcont &  0.475 $\pm$ 0.177  \\
\cutinhead{IS Tau A-B}
2008.9620  &   210.48  $\pm$   0.76  &   116.87  $\pm$   0.21  & Kcont &  0.2418 $\pm$ 0.0063  \\
           &                         &                         & L$'$  &  0.2050 $\pm$ 0.0047  \\
2013.0745  &   210.63  $\pm$   5.10  &   124.79  $\pm$   1.39  & Kcont &  0.1358 $\pm$ 0.0502  \\
\cutinhead{RX J0435.9+2352 A-B}
2008.9622  &    82.28  $\pm$   1.37  &   323.33  $\pm$   0.96  & Kcont &  0.420 $\pm$ 0.013  \\
           &                         &                         & Hcont &  0.368 $\pm$ 0.015  \\
2011.7796  &    91.54  $\pm$   0.81  &   328.51  $\pm$   0.50  & Kcont &  0.371 $\pm$ 0.016  \\
2014.0679  &    90.09  $\pm$   0.55  &   334.75  $\pm$   0.35  & Kcont &  0.401 $\pm$ 0.013  \\
           &                         &                         & Hcont &  0.339 $\pm$ 0.010  \\
\cutinhead{T Tau Sa-Sb}
2002.8294  &   107.35  $\pm$   1.76  &   283.78  $\pm$   0.94  & K$'$  &  3.3764 $\pm$ 0.1045 \\
           &                         &                         & Br$\gamma$ & 3.4498 $\pm$ 0.0903 \\
2004.9814  &   119.87  $\pm$   1.50  &   296.27  $\pm$   0.72  & Kcont &  2.9912 $\pm$ 0.1019 \\
2005.9315  &   122.21  $\pm$   5.35  &   303.80  $\pm$   2.51  & Kcont &  1.9640 $\pm$ 0.2021 \\
2006.9630  &   125.50  $\pm$   0.22  &   305.58  $\pm$   0.10  & Kcont &  0.8997 $\pm$ 0.0141 \\
2008.0449  &   127.47  $\pm$   0.17  &   310.65  $\pm$   0.08  & Kcont &  0.7862 $\pm$ 0.0067 \\
2008.9617  &   128.17  $\pm$   0.50  &   314.62  $\pm$   0.22  & Kcont &  2.0088 $\pm$ 0.0238 \\
2009.8166  &   127.65  $\pm$   0.15  &   318.93  $\pm$   0.07  & Kcont &  1.1508 $\pm$ 0.0114 \\
2011.0644  &   125.43  $\pm$   0.28  &   324.69  $\pm$   0.13  & Kcont &  0.9348 $\pm$ 0.0143 \\
           &                         &                         & Hcont &  5.6955 $\pm$ 1.0035 \\
2011.7797  &   123.81  $\pm$   0.26  &   327.83  $\pm$   0.12  & Kcont &  1.3313 $\pm$ 0.0099 \\
2013.0740  &   119.78  $\pm$   0.42  &   334.65  $\pm$   0.20  & Kcont &  1.6169 $\pm$ 0.0455 \\
2014.0077  &   116.00  $\pm$   0.39  &   339.89  $\pm$   0.19  & Kcont &  1.6492 $\pm$ 0.0255 \\
\cutinhead{T Tau N-Sa}
2002.8294  &   698.19  $\pm$   1.81  &  182.272  $\pm$  0.149  & K$'$  &  0.0220 $\pm$ 0.0012  \\
           &                         &                         & Br$\gamma$ & 0.0200 $\pm$ 0.0006  \\
2004.9814  &   698.77  $\pm$   1.38  &  183.726  $\pm$  0.113  & Kcont &  0.0217 $\pm$ 0.0007  \\
2005.9315  &   703.05  $\pm$   4.93  &  184.723  $\pm$  0.402  & Kcont &  0.0342 $\pm$ 0.0033  \\
2006.9630  &   697.18  $\pm$   0.58  &  185.113  $\pm$  0.048  & Kcont &  0.0873 $\pm$ 0.0028  \\
2008.0449  &   697.33  $\pm$   0.64  &  185.908  $\pm$  0.053  & Kcont &  0.1170 $\pm$ 0.0025  \\
2008.9617  &   696.85  $\pm$   0.87  &  186.556  $\pm$  0.072  & Kcont &  0.0470 $\pm$ 0.0009  \\
2009.8166  &   696.84  $\pm$   0.37  &  187.286  $\pm$  0.032  & Kcont &  0.0975 $\pm$ 0.0010  \\
2011.0644  &   694.98  $\pm$   0.55  &  188.279  $\pm$  0.046  & Kcont &  0.1314 $\pm$ 0.0037  \\
           &                         &                         & Hcont &  0.0035 $\pm$ 0.0006  \\
2011.7797  &   694.09  $\pm$   0.28  &  188.775  $\pm$  0.025  & Kcont &  0.0725 $\pm$ 0.0008  \\
2013.0740  &   693.23  $\pm$   0.49  &  189.907  $\pm$  0.041  & Kcont &  0.0603 $\pm$ 0.0014  \\
2014.0077  &   692.30  $\pm$   0.60  &  190.723  $\pm$  0.051  & Kcont &  0.0497 $\pm$ 0.0010  \\
\cutinhead{T Tau N-Sb}
2002.8294  &   684.90  $\pm$   0.82  &  191.107  $\pm$  0.069  & K$'$  &  0.0743 $\pm$ 0.0026  \\
           &                         &                         & Br$\gamma$ & 0.0688 $\pm$ 0.0014  \\
2004.9814  &   662.13  $\pm$   0.93  &  193.351  $\pm$  0.081  & Kcont &  0.0649 $\pm$ 0.0022  \\
2005.9315  &   652.46  $\pm$   3.18  &  194.144  $\pm$  0.279  & Kcont &  0.0667 $\pm$ 0.0040  \\
2006.9630  &   642.71  $\pm$   0.49  &  194.803  $\pm$  0.044  & Kcont &  0.0785 $\pm$ 0.0015  \\
2008.0449  &   633.40  $\pm$   0.55  &  195.426  $\pm$  0.051  & Kcont &  0.0920 $\pm$ 0.0017  \\
2008.9617  &   626.01  $\pm$   0.66  &  195.832  $\pm$  0.062  & Kcont &  0.0945 $\pm$ 0.0013  \\
2009.8166  &   619.41  $\pm$   0.42  &  196.145  $\pm$  0.040  & Kcont &  0.1122 $\pm$ 0.0008  \\
2011.0644  &   610.28  $\pm$   0.42  &  196.425  $\pm$  0.040  & Kcont &  0.1228 $\pm$ 0.0022  \\
           &                         &                         & Hcont &  0.0196 $\pm$ 0.0010  \\
2011.7797  &   606.03  $\pm$   0.24  &  196.469  $\pm$  0.024  & Kcont &  0.0965 $\pm$ 0.0012  \\
           &                         &                         & Hcont &  0.0171 $\pm$ 0.0009  \\
2013.0740  &   599.43  $\pm$   0.32  &  196.531  $\pm$  0.031  & Kcont &  0.0975 $\pm$ 0.0017  \\
           &                         &                         & Hcont &  0.0177 $\pm$ 0.0006  \\
2014.0077  &   595.66  $\pm$   0.46  &  196.451  $\pm$  0.045  & Kcont &  0.0822 $\pm$ 0.0015  \\
           &                         &                         & Hcont &: 0.0198 $\pm$ 0.0007  \\
\cutinhead{UX Tau A-Ba}
2009.8166  &  5884.25  $\pm$   2.04  &  269.495  $\pm$  0.022  & Kcont &  0.1461 $\pm$ 0.0049  \\
2011.7795  &  5892.94  $\pm$   1.01  &  269.500  $\pm$  0.013  & Kcont &  0.1602 $\pm$ 0.0067  \\
2013.0740  &  5905.41  $\pm$   3.89  &  269.593  $\pm$  0.039  & Kcont &  0.1334 $\pm$ 0.0268  \\
\cutinhead{UX Tau A-Bb}
2009.8166  &  5930.07  $\pm$   1.61  &  270.181  $\pm$  0.018  & Kcont &  0.1184 $\pm$ 0.0037  \\
2011.7795  &  5912.00  $\pm$   1.14  &  270.007  $\pm$  0.014  & Kcont &  0.1275 $\pm$ 0.0074  \\
2013.0740  &  5905.55  $\pm$   4.96  &  269.905  $\pm$  0.049  & Kcont &  0.1125 $\pm$ 0.0191  \\
\cutinhead{UX Tau Ba-Bb}
2009.8166  &    84.24  $\pm$   1.09  &   326.89  $\pm$   0.74  & Kcont &  0.810 $\pm$ 0.021  \\
2011.7795  &    55.53  $\pm$   0.64  &   339.68  $\pm$   0.66  & Kcont &  0.795 $\pm$ 0.013  \\
2013.0740  &    32.19  $\pm$   4.90  &   359.49  $\pm$   8.73  & Kcont &  0.908 $\pm$ 0.400  \\
\cutinhead{UX Tau A-C}
2009.8166  &  2711.04  $\pm$   3.36  &  181.605  $\pm$  0.072  & Kcont &  0.0519 $\pm$ 0.0051  \\
2011.7795  &  2715.80  $\pm$   1.33  &  181.559  $\pm$  0.030  & Kcont &  0.0564 $\pm$ 0.0032  \\
2013.0740  &  2720.17  $\pm$   2.25  &  181.598  $\pm$  0.048  & Kcont &  0.0728 $\pm$ 0.0122  \\
\cutinhead{V410 Tau A-C}
2005.9317  &   325.42  $\pm$   0.95  &   139.03  $\pm$   0.17  & Kcont &  0.0500 $\pm$ 0.0010  \\
2011.7796  &   333.29  $\pm$   0.34  &   141.69  $\pm$   0.06  & Kcont &  0.0548 $\pm$ 0.0009  \\
2013.0745  &   334.13  $\pm$   3.27  &   142.33  $\pm$   0.56  & Kcont &  0.0568 $\pm$ 0.0061  \\
\cutinhead{V928 Tau A-B}
2011.7796  &   235.62  $\pm$   0.12  &  289.827  $\pm$  0.031  & Kcont &  0.9751 $\pm$ 0.0058  \\
\cutinhead{ZZ Tau A-B}
2004.9815  &    61.32  $\pm$   1.35  &    74.02  $\pm$   1.26  & K$'$  &  0.672 $\pm$ 0.047  \\
2006.9632  &    66.31  $\pm$   1.21  &    61.39  $\pm$   1.05  & K$'$  &  0.594 $\pm$ 0.026  \\
           &                         &                         & H     &  0.630 $\pm$ 0.073  \\
           &                         &                         & J     &  0.735 $\pm$ 0.069  \\
2008.0446  &    66.31  $\pm$   0.49  &    54.09  $\pm$   0.42  & Kcont &  0.639 $\pm$ 0.010  \\
           &                         &                         & L$'$  &  0.540 $\pm$ 0.011  \\
2008.9621  &    67.75  $\pm$   0.55  &    48.80  $\pm$   0.47  & Kcont &  0.638 $\pm$ 0.016  \\
           &                         &                         & Hcont &  0.632 $\pm$ 0.018  \\
2011.0645  &    73.02  $\pm$   1.05  &    38.39  $\pm$   0.82  & Kcont &  0.644 $\pm$ 0.021  \\
           &                         &                         & Hcont &  0.620 $\pm$ 0.009  \\
2011.7797  &    74.10  $\pm$   0.41  &    33.89  $\pm$   0.32  & Kcont &  0.675 $\pm$ 0.027  \\
           &                         &                         & Hcont &  0.588 $\pm$ 0.034  \\
2013.0741  &    76.77  $\pm$   1.56  &    23.74  $\pm$   1.16  & Kcont &  0.619 $\pm$ 0.022  \\
2014.0679  &    79.52  $\pm$   0.63  &    23.37  $\pm$   0.46  & Kcont &  0.641 $\pm$ 0.014  \\
           &                         &                         & Hcont &  0.677 $\pm$ 0.018  \\
\enddata 
\tablenotetext{a}{For each system, the ``X-Y'' designation indicates that component Y was measured relative to X.}
\tablenotetext{b}{HIP 23418 is a member of $\beta$ Pic moving group \citep{song03}.  We observed it as a PSF star, but present our measurement of its separation, position angle, and flux ratio even though it is not part of the Taurus star forming region.}
\label{tab.sepPA}
\end{deluxetable} 

\clearpage

\begin{deluxetable}{lllllllll}
\tabletypesize{\scriptsize}
\rotate
\tablecaption{Preliminary Orbital Parameters\tablenotemark{a}}
\tablewidth{0pt}
\tablehead{
\colhead{System} & \colhead{$P$} & \colhead{$T$} & \colhead{$e$} & \colhead{$a$} & \colhead{$i$} & \colhead{$\Omega$} & \colhead{$\omega$}  & \colhead{$M_{\rm tot}~(\frac{d}{D})^3$} \\
\colhead{}       & \colhead{(y)} & \colhead{}    & \colhead{}  & \colhead{(mas)} & \colhead{($^\circ$)} & \colhead{($^\circ$)} & \colhead{($^\circ$)}  & \colhead{($M_{\odot}$)} }
\startdata
DF Tau A-B  & 43.7 $\pm$ 3.0 & 1980.5  $\pm$ 1.7  & 0.287 $\pm$ 0.067 & 93.5 $\pm$ 1.8 & 144.3 $\pm$ 2.1  & 27.1  $\pm$ 8.2  & 302.6 $\pm$ 7.6  & 1.17 $\pm$ 0.13 $\pm$ 0.25 \\ 
HBC 351 A-B\tablenotemark{b} & 30.16 $\pm$ 0.27 & 2011.949 $\pm$ 0.012 & 0.5562 $\pm$ 0.0018 & 80.84 $\pm$ 0.40 & 47.67 $\pm$ 0.34 & 22.77 $\pm$ 0.56 & 41.51 $\pm$ 0.29 & 1.42 $\pm$ 0.01 $\pm$ 0.10 \\ 
T Tau Sa-Sb & 29.0 $\pm$ 3.6 & 1995.82 $\pm$ 0.89 & 0.49  $\pm$ 0.12  & 89.0 $\pm$ 7.3 & 29.2  $\pm$ 11.8 & 107.3 $\pm$ 18.7 & 37.6  $\pm$ 14.1 & 2.70 $\pm$ 0.22 $\pm$ 0.03 \\ 
ZZ Tau A-B  & 46.2 $\pm$ 2.8 & 1995.01 $\pm$ 0.28 & 0.567 $\pm$ 0.027 & 86.4 $\pm$ 4.5 & 124.6 $\pm$ 2.0  & 137.0 $\pm$ 2.3  & 294.9 $\pm$ 2.3  & 0.83 $\pm$ 0.16 $\pm$ 0.18 \\ 
\enddata 
\tablenotetext{a}{In computing the total masses, we assumed a distance of $D=147.6 \pm 0.6$ pc for T Tau \citep{loinard07}, $D=134.6 \pm 3.1$ pc for HBC 351 \citep{soderblom05}, and the average distance of $D=140 \pm 10$ pc to the Taurus star forming region \citep{kenyon94} for DF Tau and ZZ Tau.  The first uncertainty in the total mass reflects only the uncertainties in the visual orbit.  The second uncertainty is the contribution from the distance which can be improved independently from the orbit.  The masses will scale as a ratio of their actual distance $d$ compared with the assumed distance $D$ as a function of $(d/D)^3$.}
\tablenotetext{b}{HBC 351 is likely a member of the Pleiades (see discussion in Section~\ref{sect.prelim})}
\label{tab.orbpar}
\end{deluxetable} 

\begin{deluxetable}{lcccc}
\tablecaption{Relative proper motions for systems showing linear motion\tablenotemark{a}}
\tablewidth{0pt}
\tablehead{
\colhead{System} & \colhead{$\mu_\alpha \cos{\delta}$} & \colhead{$\mu_\delta$} & \colhead{$\Delta \mu_\alpha \cos{\delta}$} & \colhead{$\Delta \mu_\delta$}  \\
\colhead{} & \colhead{(mas yr$^{-1}$)} & \colhead{(mas yr$^{-1}$)} & \colhead{(mas yr$^{-1}$)} & \colhead{(mas yr$^{-1}$)} }
\startdata
FW Tau\tablenotemark{b}     &  10 &  -22 &  -3.48 $\pm$ 0.11  &   8.73 $\pm$ 0.13  \\ 
GH Tau     &   3 &  -30 &  -0.37 $\pm$ 0.05  &  -3.17 $\pm$ 0.04  \\
Haro 6-37  &   0 &  -14 &   2.54 $\pm$ 0.27  &   1.75 $\pm$ 0.26  \\
HBC 360    & -11 &    8 &  -1.47 $\pm$ 0.12  &   3.30 $\pm$ 0.12  \\
IS Tau     &   9 &  -28 &  -2.80 $\pm$ 0.12  &  -5.45 $\pm$ 0.12  \\
V928 Tau   &   4 &  -28 &  -4.08 $\pm$ 0.07  &  -0.82 $\pm$ 0.07  \\
\enddata 
\tablenotetext{a}{$\mu_\alpha \cos{\delta}$ and $\mu_\delta$ are from the proper motion catalog of \citet{ducourant05}. $\Delta \mu_\alpha \cos{\delta}$ and $\Delta \mu_\delta$ are the relative motion of the binary components based on the fit for linear motion.}
\tablenotetext{b}{FW Tau - linear option, note that orbital option gives a lower $\chi^2$.}
\label{tab.pm}
\end{deluxetable}

\begin{deluxetable}{lccccc}
\tablecaption{Average Magnitudes for Components in DF Tau and ZZ Tau}
\tablewidth{0pt}
\tablehead{
\colhead{Star} & \colhead{$V$} & \colhead{$J$} & \colhead{$H$} & \colhead{$Ks$} & \colhead{$L$} }
\startdata
DF Tau A & 12.62 $\pm$ 0.40 & 8.876 $\pm$ 0.029 & 7.865 $\pm$  0.064 & 7.176 $\pm$ 0.046 & 6.27 $\pm$ 0.15 \\
DF Tau B & 13.51 $\pm$ 0.06 & 8.973 $\pm$ 0.029 & 8.175 $\pm$  0.082 & 7.923 $\pm$ 0.082 & 7.53 $\pm$ 0.15 \\
ZZ Tau A & 14.75 $\pm$ 0.20 & 10.093 $\pm$ 0.048 & 9.225 $\pm$ 0.036 & 8.978 $\pm$ 0.027 & 8.66 $\pm$ 0.11 \\
ZZ Tau B & 15.75 $\pm$ 0.19 & 10.428 $\pm$ 0.063 & 9.728 $\pm$ 0.044 & 9.462 $\pm$ 0.034 & 9.33 $\pm$ 0.11 \\
\enddata 
\tablecomments{$V$ is from calibrated FGS photometry \citep{schaefer03,schaefer06}.  $J$, $H$, $Ks$ magnitudes determined from combining the average of our AO flux ratios with the total magnitude from 2MASS. $L$-band magnitude determined from combining our AO flux ratio with $L$-band magnitudes reported by \citet{kenyon95}.  DF Tau also has component magnitudes in $UBVRI$ from {\it HST} photometry \citep{white01}.}
\label{tab.mag}
\end{deluxetable}

\begin{deluxetable}{llclll}
\tablecaption{Physical properties of the components in DF Tau and ZZ Tau}
\tablewidth{0pt}
\tablehead{
\colhead{Star} & \colhead{SpT} & \colhead{$T_{\rm eff}$ (K)\tablenotemark{a}} & \colhead{$A_V$ (mag)} & \colhead{$L$ $(\frac{d}{\rm 140 pc})^2$ $(L_\odot)$} & \colhead{Ref} }
\startdata
DF Tau A & M1V   & 3705 $\pm$ 145 & 0.66            & 0.556 $\pm$ 0.13  & 1 \\
DF Tau B & M1V   & 3705 $\pm$ 145 & 0.75            & 0.689 $\pm$ 0.16  & 1 \\
ZZ Tau A & M2.5V & 3488 $\pm$ 145 & 1.49 $\pm$ 0.34 & 0.411 $\pm$ 0.038 & 2 \\
ZZ Tau B & M3.5V & 3343 $\pm$ 145 & 1.24 $\pm$ 0.87 & 0.241 $\pm$ 0.055 & 2 \\
\enddata 
\tablenotetext{a}{$T_{\rm eff}$ are determined from the spectral types based on the temperature scale of \citet{luhman03}.}
\tablecomments{References: (1) \citet{hillenbrand04}; (2) This work.}
\label{tab.properties}
\end{deluxetable} 


\clearpage

\clearpage

\clearpage

\clearpage


\clearpage

\clearpage

\clearpage

\clearpage

\clearpage


\clearpage

\clearpage

\clearpage

\clearpage

\clearpage

\clearpage

\clearpage

\clearpage


\clearpage

\clearpage

\clearpage

\clearpage

\clearpage

\clearpage

\end{document}